\documentclass[DIV=12,numbers=noenddot]{scrartcl}

\usepackage{graphicx}

\usepackage{subfig}
\usepackage{float}
\usepackage[utf8]{inputenc}
\usepackage[T1]{fontenc}
\usepackage{lmodern}
\usepackage{amsmath,amssymb}
\usepackage{authblk}
\usepackage{xcolor}
\usepackage{xspace}
\usepackage[inline]{enumitem}
\usepackage{autobreak}
\usepackage{soul}
\usepackage[
  backend=bibtex,    
  bibencoding=utf8,    
  sorting=none,    
  url=true,    
  doi=false,    
  style=phys,    
  biblabel=brackets,    
  subentry=false,    
  eprint=true,    
  maxbibnames=5    
]{biblatex}
\usepackage{sectsty}
\usepackage{multirow}
\allsectionsfont{\boldmath}

\newcommand{\mytitle}{How robust are particle physics predictions in asymptotic safety?}
\newcommand{\myauthors}{Wojciech Kotlarski, Kamila Kowalska, Daniele Rizzo, Enrico Maria Sessolo}

\newcommand{\pyrate}{\texttt{PyR@TE}\xspace}

\usepackage[
  pdftitle={\mytitle}, 
  pdfauthor={\myauthors},
  pdfkeywords={},
  bookmarks=true, linktocpage, colorlinks=true, allbordercolors=white,
  allcolors=blue]{hyperref}
\usepackage{orcidlink}

\newcommand{\newc}{\newcommand*}

\long\def\begincomment#1\endcomment{%
        \begingroup\sf\baselineskip12pt#1\endgroup}

\newc{\etal}{\textrm{et al.}} 
\newc{\eg}{\textrm{e.g.}} 
\newc{\ie}{\textrm{i.e.}}
\newc{\etc}{\textrm{etc}}
\newc\vs{\textrm{vs.}}
\newc{\cl}{\textrm {C.L.}}

\newc{\ev}{\ensuremath{\,\mathrm{eV}}}
\newc{\kev}{\ensuremath{\,\mathrm{keV}}}
\newc{\mev}{\ensuremath{\,\mathrm{MeV}}}
\newc{\gev}{\ensuremath{\,\mathrm{GeV}}}
\newc{\tev}{\ensuremath{\,\mathrm{TeV}}}
\newc{\MeV}{\mev} 
\newc{\TeV}{\tev}
\newc{\invpb}{\ensuremath{/\text{pb}}}
\newc{\invfb}{\ensuremath{/\text{fb}}}
\newc\nb{\ensuremath{\,\mathrm{nb}}} \newc\pb{\ensuremath{\,\mathrm{pb}}} \newc\fb{\ensuremath{\,\mathrm{fb}}}
\newc\pc{\ensuremath{\,\mathrm{pc}}}
\newc\kpc{\ensuremath{\,\mathrm{kpc}}}
\newc\mpc{\ensuremath{\,\mathrm{Mpc}}}
\newc\ps{\ensuremath{\,\mathrm{ps}}} 
\newc\cmeter{\ensuremath{\,\mathrm{cm}}} 
\newc\meter{\ensuremath{\,\mathrm{m}}} 
\newc\kmeter{\ensuremath{\,\mathrm{km}}}
\newc\second{\ensuremath{\,\mathrm{s}}}
\newc\msecond{\ensuremath{\,\mathrm{ms}}}
\newc\nsecond{\ensuremath{\,\mathrm{ns}}}
\newc\psecond{\ensuremath{\,\mathrm{ps}}}

\newc{\chisqmin}{\ensuremath{\chi^2_{\mathrm{min}}}}
\newc{\Delchisq}{\ensuremath{\Delta\chi^2}}
\newc{\chisq}{\ensuremath{\chi^2}}
\newc{\like}{\ensuremath{\mathcal{L}}}

\newc\lsim{\ensuremath{\mathrel{\rlap{\lower4pt\hbox{\hskip1pt$\sim$}}\raise1pt\hbox{$<$}}}}
\newc\gsim{\ensuremath{\mathrel{\rlap{\lower4pt\hbox{\hskip1pt$\sim$}}\raise1pt\hbox{$>$}}}}
\newc{\VEV}[1]{\ensuremath{\langle #1 \rangle}}
\newc{\dl}{\ensuremath{\stackrel{\leftarrow}{D}}}
\newc{\dr}{\ensuremath{\stackrel{\rightarrow}{D}}}

\newc{\bcenter}{\begin{center}}    \newc{\ecenter}{\end{center}}
\newc{\bfl}{\begin{flushleft}}    \newc{\efl}{\end{flushleft}}
\newc{\bfr}{\begin{flushright}}    \newc{\efr}{\end{flushright}}

\newc{\bi}{\begin{itemize}}
\newc{\ei}{\end{itemize}}
\newc{\bed}{\begin{description}}
\newc{\eed}{\end{description}}
\newc{\ben}{\begin{enumerate}}
\newc{\een}{\end{enumerate}}

\newc{\be}{\begin{equation}}
\newc{\ee}{\end{equation}}
\newc{\bea}{\begin{eqnarray}}
\newc{\eea}{\end{eqnarray}}
\newc{\bfle}{\begin{flalign}}
\newc{\efle}{\end{flalign}}
\newc{\ra}{\rightarrow}

\newc{\alphas}{\ensuremath{\alpha_s}}
\newc{\alphatwo}{\ensuremath{\alpha_2}}
\newc{\alphaone}{\ensuremath{\alpha_1}}
\newc{\alphai}[1]{\ensuremath{\alpha_{#1}}}
\newc{\alphaem}{\ensuremath{\alpha_{\mathrm{em}}}}
\newc{\alphaeff}{\ensuremath{\alpha_{\mathrm{eff}}}}
\newc{\sineff}{\ensuremath{\sin \theta_{\mathrm{eff}}}}
\newc{\sinsqeff}{\ensuremath{\sin^2 \theta_{\mathrm{eff}}}}
\newc{\dalphahad}{\ensuremath{\Delta \alpha_{\mathrm{had}}}}
\newc{\yt}{\ensuremath{h_t}} \newc{\yb}{\ensuremath{h_b}} \newc{\ytau}{\ensuremath{h_{\tau}}}
\newc\mz{\ensuremath{M_Z}} 
\newc\mw{\ensuremath{m_W}}
\newc\mZ{\mz}        \newc\mW{\mw}
\newc\mhsm{\ensuremath{ m_{H_{\mathrm{SM}}}}}
\newc{\mtop}{\ensuremath{ m_t}}               \newc{\mtpole}{\ensuremath{ M_t}}
\newc{\mbottom}{\ensuremath{ m_b}} 
\newc{\mtau}{\ensuremath{ m_{\tau}}}
\newc{\mt}{\mtpole}
\newc{\mb}{\mbottom} 
\newc{\rtwogg}{\ensuremath{R_{h_2}(\gamma\gamma)}}
\newc{\rtwozz}{\ensuremath{R_{h_2}(ZZ)}}
\newc{\ronegg}{\ensuremath{R_{h_1}(\gamma\gamma)}}
\newc{\ronezz}{\ensuremath{R_{h_1}(ZZ)}}
\newc{\rsiggg}{\ensuremath{R_{h_\textrm{sig}}(\gamma\gamma)}}
\newc{\rsigzz}{\ensuremath{R_{h_\textrm{sig}}(ZZ)}}
\newc{\llbar}{\ensuremath{\ell\bar{\ell}}}
\newc{\tauptaum}{\ensuremath{ \tau^+\tau^-}}
\newc{\qqbar}{\ensuremath{ q\bar{q}}} \newc{\ppbar}{\ensuremath{ p\bar{p}}}
\newc{\bbbar}{\ensuremath{ b\bar{b}}} \newc{\ttbar}{\ensuremath{ t\bar{t}}}
\newc{\ffbar}{\ensuremath{ f\bar{f}}} \newc{\tautaubar}{\ensuremath{ \tau\bar{\tau}}}

\newc{\mchi}{\ensuremath{m_\neutone}}
\newc{\squark}{\ensuremath{\tilde{q}}}
\newc{\slepton}{\ensuremath{\tilde{l}}}
\newc{\gluino}{\ensuremath{\tilde{g}}} 
\newc{\mgluino}{\ensuremath{{m_{\gluino}}}}
\newc{\wino}{\ensuremath{\tilde{W}}} 
\newc{\mwino}{\ensuremath{{m_{\wino}}}}
\newc{\tone}{\ensuremath{{\tilde{t}_1}}}
\newc{\Hone}{\ensuremath{{\tilde{H}_{1}}}}
\newc{\Htwo}{\ensuremath{{\tilde{H}_{2}}}}
\newc{\Hhtwo}{\ensuremath{{H_{2}}}}
\newc{\qli}{\ensuremath{{\tilde{Q}_{i}}}}
\newc{\uri}{\ensuremath{{\tilde{u}_{i}}}}
\newc{\dri}{\ensuremath{{\tilde{d}_{i}}}}
\newc{\lli}{\ensuremath{{\tilde{L}_{i}}}}
\newc{\eri}{\ensuremath{{\tilde{e}_{i}}}}

\newc{\sthw}{\ensuremath{ \sin\theta_W}}              \newc{\cthw}{\ensuremath{\cos\theta_W}}
\newc{\tanthw}{\ensuremath{ \tan\theta_W}}              \newc{\cotthw}{\ensuremath{\cot\theta_W}}
\newc{\ssqthw}{\ensuremath{\sin^2 \theta_W}}
\newc{\msbar}{\ensuremath{\overline{MS}}} \newc{\drbar}{\ensuremath{\overline{DR}}}
\newc{\mtmtsmmsbar}{\ensuremath{ m_t(m_t)^{\msbar}_{{\mathrm{SM}}}}}
\newc{\mtmtsmdrbar}{\ensuremath{ m_t(m_t)^{\drbar}_{{\mathrm{SM}}}}}
\newc{\mtmtmssmdrbar}{\ensuremath{ m_t(m_t)^{\drbar}_{{\mathrm{SUSY}}}}}
\newc{\mbmbmsbar}{\ensuremath{ m_b(m_b)^{\msbar} }}
\newc{\mbmbsmmsbar}{\ensuremath{ m_b(m_b)^{\msbar}_{{\mathrm{SM}}}}}
\newc{\mbmzsmmsbar}{\ensuremath{ m_b(\mz)^{\msbar}_{{\mathrm{SM}}}}}
\newc{\mbmzsmdrbar}{\ensuremath{ m_b(\mz)^{\drbar}_{{\mathrm{SM}}}}}
\newc{\mbmzmssmdrbar}{\ensuremath{ m_b(\mz)^{\drbar}_{{\mathrm{SUSY}}}}}
\newc{\mtaumzsmmsbar}{\ensuremath{ m_{\tau}(\mz)^{\msbar}_{{\mathrm{SM}}}}}
\newc{\mtaumzsmdrbar}{\ensuremath{ m_{\tau}(\mz)^{\drbar}_{{\mathrm{SM}}}}}
\newc{\mtaumzmssmdrbar}{\ensuremath{ m_{\tau}(\mz)^{\drbar}_{{\mathrm{SUSY}}}}}
\newc{\alphasmzms}{\ensuremath{\alpha_s(M_Z)^{\overline{MS}}}}
\newc{\alphaimzms}[1]{\ensuremath{\alpha_{#1}(M_Z)^{\overline{MS}}}}
\newc{\alphaemmz}{\ensuremath{\alpha_{\mathrm{em}}(M_Z)^{\overline{MS}}}}

\newc{\mzero}{\ensuremath{{m_0}}}
\newc{\mhalf}{\ensuremath{ m_{1/2}}}
\newc{\tanb}{\ensuremath{\tan\beta}}
\newc{\azero}{\ensuremath{ A_0}}
\newc{\signmu}{\ensuremath{\rm{sgn}\,\mu}}
\newc{\atau}{\ensuremath{{A_{\tau}}}}
\newc{\mueff}{\ensuremath{\mu_{\rm{eff}}}}
\newc{\lam}{\ensuremath{{\lambda}}}
\newc{\kap}{\ensuremath{{\kappa}}}
\newc{\alam}{\ensuremath{{A_{\lambda}}}}
\newc{\akap}{\ensuremath{{A_{\kappa}}}}
\newc{\hs}{\ensuremath{ H_s}}      
\newc{\mhs}{\ensuremath{ m_{H_s}}} 
\newc{\mgut}{\ensuremath{ M_{\rm GUT}}}
\newc{\mvl}{\ensuremath{ M_{\rm VL}}}
\newc{\gut}{\ensuremath{{\rm GUT}}}
\newc{\mplanck}{\ensuremath{ M_{\textrm P}}}      \newc{\mpl}{\ensuremath{ M_{\textrm{Pl}}}}
\newc{\msusy}{\ensuremath{ M_{\rm SUSY}}}      \newc{\ms}{\ensuremath{ M_{\rm S}}}
 \newc{\hu}{\ensuremath{ H_u}}       \newc{\hd}{\ensuremath{ H_d}}
 \newc{\mhu}{\ensuremath{ m_{H_u}}}       \newc{\mhd}{\ensuremath{ m_{H_d}}}
 \newc{\mhuew}{\ensuremath{ m^{\ast}_{H_u}}}       \newc{\mhdew}{\ensuremath{ m^{\ast}_{H_d}}}
 \newc{\mhuewsq}{\ensuremath{ m^{\ast\, 2}_{H_u}}}       \newc{\mhdewsq}{\ensuremath{ m^{\ast\, 2}_{H_d}}}
 \newc{\mhl}{\ensuremath{m_\hl}} 
 \newc{\mhone}{\ensuremath{m_{h_1}}} 
 \newc{\mhtwo}{\ensuremath{m_{h_2}}} 
 \newc{\mhi}{\ensuremath{m_{\tilde{h}}}} 
 \newc{\mul}{\ensuremath{m_{\tilde{u}_L}}} 
 \newc{\mtone}{\ensuremath{m_{\tilde{t}_1}}} 
 \newc{\ma}{\ensuremath{m_A}} 
 \newc{\mH}{\ensuremath{m_H}} 
 \newc{\maone}{\ensuremath{m_{a_1}}} 
 \newc{\matwo}{\ensuremath{m_{a_2}}}
 \newc{\hone}{\ensuremath{h_1}}
 \newc{\htwo}{\ensuremath{h_2}}
 \newc{\aone}{\ensuremath{a_1}}
 \newc{\atwo}{\ensuremath{a_2}}
 \newc{\mqthree}{\ensuremath{m_{\tilde{Q}_3}^2}}
 \newc{\muthree}{\ensuremath{m_{\tilde{u}_3}^2}}
 \newc{\mqli}{\ensuremath{m_{\tilde{Q}_{i}}}}
 \newc{\muri}{\ensuremath{m_{\tilde{u}_{i}}}}
 \newc{\mdri}{\ensuremath{m_{\tilde{d}_{i}}}}
 \newc{\mlli}{\ensuremath{m_{\tilde{L}_{i}}}}
 \newc{\meri}{\ensuremath{m_{\tilde{e}_{i}}}}
 \newc{\ts}{\ensuremath{T_{SUSY}}}

\newc{\sigsip}{\ensuremath{\sigma^{\rm SI}_{p}}}	\newc{\sigsin}{\ensuremath{\sigma^{\rm SI}_{n}}}
\newc{\sigsdp}{\ensuremath{\sigma^{\rm SD}_{p}}}	\newc{\sigsdn}{\ensuremath{\sigma^{\rm SD}_{n}}}
\newc{\sigsi}{\ensuremath{\sigma^{\rm SI}}}	\newc{\sigsd}{\ensuremath{\sigma^{\rm SD}}}
\newc{\abund}{\ensuremath{ \Omega h^2}}
\newc{\omegadm}{\ensuremath{ \Omega_{{\rm DM}}}}     \newc{\abunddm}{\ensuremath{ \Omega_{{\rm DM}} h^2}} 
\newc{\omegam}{\ensuremath{ \Omega_{{\rm m}}}}       \newc{\abundm}{\ensuremath{ \Omega_{{\rm m}} h^2}}
\newc{\omegab}{\ensuremath{ \Omega_{{\rm b}}}}	\newc{\abundb}{\ensuremath{ \Omega_{{\rm b}} h^2}}
\newc{\omegatot}{\ensuremath{ \Omega_{{\rm TOT}}}}
\newc{\omegacdm}{\ensuremath{ \Omega_{{\rm CDM}}}}   \newc{\abundcdm}{\ensuremath{ \Omega_{{\rm CDM}} h^2}}
\newc{\omegalambda}{\ensuremath{ \Omega_{\Lambda}}} \newc{\abundlambda}{\ensuremath{ \Omega_{\Lambda} h^2}}
\newc{\omegarad}{\ensuremath{ \Omega_{{\rm rad}}}}  \newc{\abundrad}{\ensuremath{ \Omega_{{\rm rad}} h^2}}
\newc{\rhocrit}{\ensuremath{ \rho_{\rm crit}}}
\newc{\rhochi}{\ensuremath{ \rho_{\chi}}}
\newc{\abunchi}{\ensuremath{\Omega_\chi h^2}}
\newc{\abundlsp}{\ensuremath{\Omega_{\rm LSP}h^2}}

\newc{\amu}{\ensuremath{ a_{\mu}}}        \newc{\amususy}{\ensuremath{ a_{\mu}^{\mathrm{SUSY}}}}
\newc{\amuexpt}{\ensuremath{ a_{\mu}^{\mathrm{expt}}}}        \newc{\amusm}{\ensuremath{ a_{\mu}^{\mathrm{SM}}}}
\newc\deltaamu{\ensuremath{\Delta a_{\mu}}} \newc{\deltaamususy}{\ensuremath{\delta a_{\mu}^{\mathrm{SUSY}}}}
\newc\gmtwo{\ensuremath{ (g-2)_{\mu}}} 
\newc{\deltagmtwomususy}{\ensuremath{\delta\left(g-2\right)_{\mu}^{\mathrm{SUSY}}}}
\newc{\deltagmtwomu}{\ensuremath{\delta\left(g-2\right)_{\mu}}}
\newc\BR{\ensuremath{\rm BR}}
\newc\bsgamma{\ensuremath{ b\rightarrow s \gamma }}
\newc\bxsgamma{\ensuremath{\overline{B}\rightarrow X_{s}\gamma}}
\newc\brbsgamma{\ensuremath{\BR\left(\bsgamma\right)}}
\newc\brbxsgamma{\ensuremath{\BR\left(\bxsgamma\right)}}
\newc\bsmumu{\ensuremath{B_s\to\mu^+\mu^-}}
\newc\brbsmumu{\ensuremath{\BR\left(B_s\to\mu^+\mu^-\right)}}
\newc\bdmmumu{\ensuremath{\overline{B}_d\to\mu^+\mu^-}}
\newc\bbbarmix{\ensuremath{\overline{B}_s\mbox{-}B_s}}      
\newc\delmbs{\ensuremath{\Delta M_{B_s}}}
\newc{\butaunu}{\ensuremath{B_u \rightarrow \tau \nu}}
\newc{\brbutaunu}{\ensuremath{\BR\left(B_u \rightarrow \tau \nu\right)}}

\newcommand*{\reftable}[1]{Table~\ref{#1}}         
       
\newcommand*{\reffig}[1]{Fig.~\ref{#1}}

        \newcommand*{\refeq}[1]{Eq.~(\ref{#1})}   

\newcommand*{\neutone}{\ensuremath{\tilde{\chi}^0_1}}

\let\oldcite\cite
\renewcommand*{\cite}{~\oldcite}

\title{\mytitle}
\author{Wojciech Kotlarski\orcidlink{0000-0002-1191-6343}\thanks{\href{mailto:wojciech.kotlarski@ncbj.gov.pl}{wojciech.kotlarski@ncbj.gov.pl}} }
\author{Kamila Kowalska\orcidlink{0000-0001-8457-2847}\thanks{\href{mailto:kamila.kowalska@ncbj.gov.pl}{kamila.kowalska@ncbj.gov.pl}} }
\author{Daniele Rizzo\orcidlink{0000-0002-2255-5272}\thanks{\href{mailto:daniele.rizzo@ncbj.gov.pl}{daniele.rizzo@ncbj.gov.pl}} }
\author{Enrico Maria Sessolo\orcidlink{0000-0001-6571-6382}\thanks{\href{mailto:enrico.sessolo@ncbj.gov.pl}{enrico.sessolo@ncbj.gov.pl}}}
\affil{National Centre for Nuclear Research \\ Pasteura 7, 02-093 Warsaw, Poland}

\date{}

\begin{document}

\maketitle
\begin{abstract}
The framework of trans-Planckian asymptotic safety has been shown to generate phenomenological predictions in the Standard Model and in some of its simple new physics extensions. A heuristic approach is often adopted, which bypasses the functional renormalization group by relying on a parametric description 
of quantum gravity with universal coefficients 
that are eventually obtained from low-energy observations. 
Within this approach a few simplifying approximations are typically introduced, including the computation of matter renormalization group equations at 1~loop, an arbitrary definition of the position of the Planck scale at $10^{19}\gev$, and an instantaneous decoupling of gravitational interactions below the Planck scale. 
In this work we systematically investigate, both analytically and numerically, the impact of dropping each of those approximations on the predictions for certain  particle physics scenarios. In particular we study two extensions of the Standard Model, the gauged $B-L$ model and the leptoquark $S_3$ model, for which we determine a set of irrelevant gauge and Yukawa couplings.  In each model, we present numerical and analytical estimates of the uncertainties associated with the predictions from asymptotic safety.
\end{abstract}

\newpage
\tableofcontents
\clearpage

\section{Introduction}
\label{sec:intro}

Asymptotic safety (AS) is the property of a 
quantum field theory to develop ultraviolet~(UV)
fixed points of the renormalization group (RG) flow of the action\cite{inbookWS}. Following the development of functional renormalization group (FRG) techniques three decades ago\cite{WETTERICH199390,Morris:1993qb}, it was shown in numerous papers that AS may arise quite naturally in quantum gravity and provide the key ingredient for the non-perturbative renormalizability of the theory. Fixed points for the rescaled Newton coupling and the cosmological constant were identified initially in the Einstein-Hilbert truncation of the effective action\cite{Reuter:1996cp,Lauscher:2001ya,Reuter:2001ag}, 
based on two operators, and later confirmed in the presence of gravitational operators of increasing mass dimension\cite{Lauscher:2002sq,Litim:2003vp,Codello:2006in,Machado:2007ea,Codello:2008vh,Benedetti:2009rx,Dietz:2012ic,Falls:2013bv,Falls:2014tra}, and of matter-field operators\cite{Oda:2015sma,Hamada:2017rvn,Christiansen:2017cxa}. 

From the point of view of particle physics in four space-time dimensions,
a particularly exciting possibility is that not only the gravitational action but the full system of gravity and matter may feature UV fixed points in the energy regime where gravitational interactions become strong\cite{Robinson:2005fj,Pietrykowski:2006xy,Toms:2007sk,Tang:2008ah,Toms:2008dq,Rodigast:2009zj,Zanusso:2009bs,Daum:2009dn,Daum:2010bc,Folkerts:2011jz,Eichhorn:2016esv,Eichhorn:2017eht}. 
A trans-Planckian fixed point may provide in that case 
specific boundary conditions for some of the \emph{a priori} free couplings of the matter Lagrangian, 
as long as they correspond to \textit{irrelevant} directions in theory space.
In this context, indications of AS emerging in U(1) gauge theories have led to discovering a gravity-driven solution to the triviality problem of the Standard Model~(SM) hypercharge coupling\cite{Harst:2011zx,Christiansen:2017gtg,Eichhorn:2017lry}. Early achievements of embedding realistic matter systems in AS include the ballpark prediction for the value of the Higgs mass (more precisely, of the quartic coupling of the Higgs potential) obtained a few 
years ahead of its discovery\cite{Shaposhnikov:2009pv} (see also Refs.\cite{Eichhorn:2017als,Kwapisz:2019wrl,Eichhorn:2021tsx}), and the retroactive ``postdiction'' of the top-mass
value\cite{Eichhorn:2017ylw}.

In order to properly complete a matter system with trans-Planckian AS, one should consistently calculate gravitational corrections to the matter beta function using the formalism of the FRG. The ensuing high-scale boundary conditions for the matter couplings, obtained this way from first principles, should be then compared with observations (cross sections, decay rates, etc.) after running the couplings to the low scale.  
It has been long known, however, that calculations based on the FRG can be subject to large theoretical uncertainties, stemming from a variety of sources -- from the choice of truncation in the gravity action\cite{Lauscher:2002sq,Codello:2007bd,Benedetti:2009rx,Falls:2017lst,Falls:2018ylp}, to cutoff-scheme dependence\cite{Reuter:2001ag,Narain:2009qa}, to the backreaction of matter, which may introduce a dependence of the gravitational fixed point on the specifics of the matter sector\cite{Percacci:2002ie,Percacci:2003jz,Dona:2013qba}. Various results can differ by up to several times\cite{Dona:2013qba} so that, often, a precise quantitative determination of the high-scale value of the matter couplings is not possible.  
Moreover, it has been recently pointed out\cite{Donoghue:2019clr,Bonanno:2020bil} that  some conceptual questions regarding the appropriateness of the FRG itself as the right tool for investigating/enforcing AS in quantum gravity remain open.

On the other hand, because of the universal nature of trans-Planckian interactions, and thanks to the ``rigidity'' 
of the dynamics in the vicinity of an interactive fixed point -- by virtue of which all quantities can be predicted except for a handful of relevant parameters that will have to be determined experimentally -- a first-principle calculation of the gravitational contribution to the matter couplings is not necessarily needed to prove the consistency of certain low-energy predictions with quantum gravity. Often one is content with establishing a heuristic framework in which the trans-Planckian interactions are parameterized by coefficients 
that are eventually obtained from low-energy observations\cite{Eichhorn:2018whv,Grabowski:2018fjj,Reichert:2019car,Alkofer:2020vtb,Kowalska:2020gie,Kowalska:2020zve,Kowalska:2022ypk,Chikkaballi:2022urc,Boos:2022jvc,Boos:2022pyq,Eichhorn:2022vgp}.

In such an effective trans-Planckian embedding, one generally introduces parametric corrections to the  
renormalization group equations (RGEs) of the renormalizable matter couplings, which take the form  
\bea
\frac{d g_i}{d t}&=&\beta_i^{\textrm{matter}}-f_g\, g_i \label{eq:betag} \\
\frac{d y_j}{d t}&=&\beta_j^{\textrm{matter}}-f_y\, y_j \label{eq:betay}\,,
\eea
where $t=\ln\mu$ (renormalization scale),
$g_i$ and $y_j$ (with $i,j=1,2,\dots$) indicate, respectively, 
the set of gauge and Yukawa couplings of the theory, and  
$\beta_{i,j}^{\textrm{matter}}$ are the beta functions of the matter theory, which can be evaluated at 1 loop in dimensional regularization~(DREG).\footnote{As long as we are interested in renormalizable couplings of the matter theory the form of the 1-loop beta function obtained in DREG coincides with the FRG calculation\cite{Baldazzi:2020vxk}.}
The two ``gravitational'' coefficients $f_g$ and $f_y$ multiply linearly all matter couplings of the same kind. They are thus universal, in the sense that gravity is not expected to be affected by the internal degrees of freedom of the matter system. They appear only in the regime where the gravitational action develops an interactive fixed point, at $\mu > M_{\textrm{Pl}}$, and serve the purpose of inducing trans-Planckian zeros on the matter beta functions. If some of the emerging fixed-point coupling values correspond to irrelevant directions of the RG flow, one can estimate $f_g$ and $f_y$ by requiring that the irrelevant fixed points should be connected, along a unique RG trajectory, to quantities measured in experiments at the low scale. 

Such heuristic embedding of a gauge-Yukawa system in trans-Planckian AS has been used 
in the SM to attempt a prediction of the
top/bottom mass ratio\cite{Eichhorn:2018whv}, the Cabibbo-Kobayashi-Maskawa\cite{Alkofer:2020vtb}, and Pontecorvo-Maki-Nakagawa-Sakata\cite{Kowalska:2022ypk} matrix elements. 
In the context of neutrino mass generation, this effective approach was employed to theorize the interesting possibility that the tiny Yukawa couplings of Dirac-type right-handed neutrinos 
may find their origin in the dynamics of the trans-Planckian flow\cite{Kowalska:2022ypk,Eichhorn:2022vgp}. 
New physics~(NP) predictions were extracted for 
neutrino masses\cite{Grabowski:2018fjj}, leptoquarks\cite{Kowalska:2020gie}, 
vector-like fermions\cite{Kowalska:2020zve}, and $\gamma/Z'$ kinetic mixing\cite{Chikkaballi:2022urc}. 
An expression similar to Eqs.~(\ref{eq:betag}) and (\ref{eq:betay}) 
potentially applies to the quartic couplings of the scalar potential as well, 
to which beta functions one could in principle assign a universal correction factor $f_{\lam}$. Following this, predictions were made for the relic abundance of dark matter\cite{Reichert:2019car,Eichhorn:2020kca}, baryon number\cite{Boos:2022jvc,Boos:2022pyq}, as well as axion models\cite{deBrito:2021akp}.\footnote{Note, however,  that whether gravitational corrections to the running couplings of the scalar potential are multiplicative or not remains a model-dependent issue, as some truncations of the gravitational action can generate additive contributions to the matter beta functions of the scalar potential\cite{Eichhorn:2020kca}.}

While in the absence of a fully developed theory of quantum gravity the heuristic approach described above has proven to be extremely fruitful for phenomenological studies in particle physics, it is also important to be aware that it is 
based on several simplifying approximations: 
\begin{itemize}
\item The DREG matter beta functions are typically computed at 1-loop
\item The Planck scale is set arbitrarily at $M_{\textrm{Pl}}=10^{19}\gev$
\item The scale dependence of $f_g$ and $f_y$, which should parameterize the cross-over from the interactive to non-interactive regime of quantum gravity, is neglected.  $f_g$ and $f_y$ are treated as constants above the Planck scale and are set to zero below. In other words, gravity decouples instantaneously at $M_{\textrm{Pl}}=10^{19}\gev$.
\end{itemize}
The question then naturally arises as to how robust the predictions 
derived in this way can be considered, and to what extent dropping any of the approximations listed above may affect a potential observational strategy to test these predictions at the low scale.  

In this study we attempt to address the issue in a systematic way. We analyze the effects of discarding one by one the approximations 
of the minimal parametric setup. Specifically we consider 
\begin{itemize}
\item the inclusion of higher-order corrections in the matter sector
\item changing the position of the Planck scale by a few orders of magnitude
\item the non-trivial functional dependence of the running gravitational couplings, 
$f_{g,y}(t)$, resulting in the non-instantaneous decoupling of the trans-Planckian UV completion.
\end{itemize}

We limit ourselves to the study of gauge-Yukawa systems of the type~(\ref{eq:betag}) and (\ref{eq:betay}).  As we have mentioned above, 
the dimensionless parameters of the scalar potential lie on a 
slightly different footing. Besides the issue of whether the gravitational correction is multiplicative or not, it has also been
shown that in the SM\cite{Kwapisz:2019wrl} and some models of NP\cite{Chikkaballi:2022urc} it is difficult to obtain the precise value of the Higgs mass if the flow originates from a UV irrelevant fixed point, so that the predictivity of AS for the scalar potential is often in question and a model-dependent issue. On the other hand, 
from the point of view of the RG flow, the parameters of the scalar potential result somewhat ``decoupled'' from the gauge-Yukawa system, as they can only affect
\refeq{eq:betag}  from the third-loop level up, and \refeq{eq:betay} from the second loop. We shall quantify, when necessary, the size of the corrections that unknown parameters of the scalar potential induce on the gauge-Yukawa system via higher-order contributions.  

Our main finding is that the predictions in the gauge sector are extremely robust. Each and every one of the effects listed above 
induce an uncertainty that does not exceed the one percent level. The situation is slightly more intricate in the Yukawa sector, where the uncertainty strongly depends on the size of the predicted coupling itself. We find that, while Yukawas as large as the SM hypercharge coupling are subject to uncertainties not exceeding the five percent level, 
if they turn out to be much smaller --
as a result of fine tuning -- their uncertainties increase significantly.

The paper is organized as follows.
In Sec.~\ref{sec:predictions} we introduce some general notions about the heuristic approach to AS, which we then use to derive benchmark predictions for two NP models. In Sec.~\ref{sec:uncertainty} we investigate the impact of abandoning one by one the simplifying approximations listed above and we derive analytical and numerical estimates of the resulting uncertainties. We present our conclusions in Sec.~\ref{sec:summary}. We give the explicit form of the 1- and 2-loop RGEs used in this work in Appendix~\ref{app:rges}.
\section{Reference model predictions at 1 loop}
\label{sec:predictions}

In order to quantify the theoretical uncertainties originating from the approximations 
listed in Sec.~\ref{sec:intro}, 
one needs a set of ``standard candle'' predictions to observe how they are modified when those approximations are dropped. 
In this section we introduce two popular renormalizable NP scenarios as reference models and we derive, in the approximations
of Sec.~\ref{sec:intro},  
the AS-driven predictions for the couplings
of their Lagrangian. In Sec.~\ref{sec:uncertainty}, 
we will drop the approximations one by one and quantify the relative change 
in the predicted values. 

\subsection{General notions\label{sec:gen_not}}

We consider typical extensions of the SM characterized 
by a set of new abelian gauge couplings and new Yukawa couplings. We compute at 1~loop the beta functions 
of the full, extended gauge-Yukawa system, and we correct the RGEs at $\mu\geq M_{\textrm{Pl}} $ linearly in the couplings,
like in Eqs.~(\ref{eq:betag}) and (\ref{eq:betay}). 

Besides the fundamental assumptions of universality 
and linearity in the matter couplings,  
the gravity parameters $f_g$ and $f_y$ 
must satisfy some loose requirements 
to be in agreement with 
the current knowledge of explicit FRG calculations. 
In particular, it has long been known that $f_g$ is likely not negative, irrespective of the chosen RG scheme\cite{Folkerts:2011jz}. More specifically, imposing $f_g>0$ will enforce asymptotic freedom 
in the non-abelian gauge sector of the SM. 
Conversely, no essential constraints 
apply to the leading-order gravitational term $f_y$.
The gravitational contribution to the Yukawa coupling was investigated in a set of simplified models\cite{Rodigast:2009zj,Zanusso:2009bs,Oda:2015sma,Eichhorn:2016esv}, 
but no general results and definite conclusions regarding the size and sign of $f_y$ are available. 

A fixed point of the system is defined by any set $\{g_i^\ast,y_j^\ast\}$, 
generically denoted with an asterisk,
for which the beta functions develop a zero: $\beta_{i(j)}^{\textrm{matter}}(g_i^\ast,y_j^\ast)-f_{g(y)}\,g_i^{\ast}(y_j^{\ast})=0$.
The RGEs of couplings $\{\alpha_k\}\equiv\{g_i,y_j\}$ are subsequently linearized
around the fixed point to derive the stability matrix $M_{ij}$\,, which is defined as
\be\label{stab}
M_{ij}=\partial\beta_i/\partial\alpha_j|_{\{\alpha^{\ast}_i\}}\,.
\ee
Eigenvalues of the stability matrix define the opposite of critical exponents $\theta_i$, 
which characterize the power-law evolution of the couplings in the vicinity of the fixed point. 
If $\theta_i$ is positive, the corresponding eigendirection is dubbed as \textit{relevant} and UV-attractive. All RG trajectories along this direction will asymptotically reach the fixed point and, 
as a consequence, a deviation of a relevant coupling from the fixed point introduces a free parameter in the theory (this freedom can be used to adjust the coupling at the high scale so that it matches 
an eventual measurement at the low scale). If $\theta_i$ is negative, 
the corresponding eigendirection is dubbed as \textit{irrelevant} and UV-repulsive. 
As we have stated already, there exists in this case only one trajectory that the coupling's flow can follow in its run to the low scale, thus potentially providing a clear prediction for its value at the experimentally accessible scale. Finally, $\theta_i=0$ corresponds to a \textit{marginal} eigendirection. The RG flow along this direction is logarithmically slow and one ought to go beyond the linear approximation provided by the stability matrix 
to decide whether a fixed point is UV-attractive or UV-repulsive.

While the approximations listed in Sec.~\ref{sec:intro} can be abandoned to evaluate the precision of the obtained predictions, the whole heuristic approach relies 
on a few assumptions which cannot be removed without compromising critically the predictive power of the fixed point analysis:
\begin{enumerate}
\item The trans-Planckian UV completion (whether it be gravity of else) 
should be responsible for the disappearance of the Landau pole in the hypercharge gauge coupling of the SM and for the appearance of an irrelevant fixed point in its place
\item The trans-Planckian UV completion should be responsible for the appearance of an irrelevant fixed point in at least one of the SM Yukawa couplings
\item The RGEs of the NP model completed in the trans-Planckian UV should not be altered by large couplings below the Planck scale. Additional light or heavy states can actually exist, but they ought to be characterized by feeble interactions with the visible sector -- see, \textit{e.g.}, Appendix~D in Ref.\cite{Chikkaballi:2022urc}.
\end{enumerate}
Point~1 and 3 allow one to extract the value of $f_g$ uniquely, by connecting
the hypercharge gauge coupling flow from the Planck scale down to the electroweak symmetry breaking (EWSB) scale. 
Point~1,
together with the fact that $f_g>0$, implies that the non-abelian gauge couplings of the SM,
$g_3$ and $g_2$, remain relevant and asymptotically free. Point~2 implies that one can connect 
an irrelevant SM direction from the Planck to the EWSB scale and 
extract uniquely the trans-Planckian value of $f_y$. All the NP Yukawa couplings for which we seek a prediction should 
correspond to irrelevant directions, while the SM Yukawa couplings that are not employed for the determination 
of $f_y$, as well as other possible couplings (point~3), 
will vanish in the deep UV and correspond to relevant directions in the coupling space. 

We emphasize that the predictions from the trans-Planckian fixed point analysis are extracted for specific NP models characterized by specific symmetries and particle content, whose RGEs should not be altered significantly below the Planck scale (point 3). As the obtained predictions are typically rather precise, this assumption, of quantum gravity being the only allowed strongly interacting UV completion to the chosen model, can be experimentally falsified. Any other strongly (or weakly) interactive UV completion defined below the Planck scale would have to be treated as a different model altogether, giving rise to different predictions and experimental signatures.

We introduce in the rest of this section two models: the $B-L$ model, 
for which we seek to predict the dark abelian gauge coupling, the kinetic mixing, and the NP Yukawa couplings associated with the right-handed neutrino sector; and 
a leptoquark~(LQ) model, for which we seek to predict the NP Yukawa interaction of the color-charged 
leptoquark with a SM quark and a lepton. For each of these models we derive the RGEs using \pyrate\,\texttt{3}\cite{Poole:2019kcm,Sartore:2020gou}.\footnote{We use the most recent available version of \pyrate  (\href{https://github.com/LSartore/pyrate/tree/master}{\pyrate master branch}). \pyrate model file for the $S_3$ model is attached to the \texttt{arXiv} version of this paper. For the gauged $B-L$ we use the model file distributed with \pyrate.}
The explicit forms of the beta functions for the models considered in this study are given in Appendix~\ref{app:rges}.

\subsection{Gauged $B-L$\label{sec:gBmL}} 

We derive predictions for the NP sector 
of the gauged $B-L$ model\cite{Coriano:2015sea,Lyonnet:2016xiz}. 
The SM symmetry is extended by an abelian gauge group U(1)$_{B-L}$, with gauge coupling $g_{B-L}$. 
The particle content of the SM is extended by one or more right-handed neutrinos and a complex scalar field 
$S$, whose vacuum expectation value~(vev) spontaneously breaks U(1)$_{B-L}$. The abelian charges of the SM and NP fields can be found, \textit{e.g.}, in Refs.\cite{Coriano:2015sea,Lyonnet:2016xiz}. 

The Yukawa part of the SM Lagrangian is extended by terms
\be
\mathcal{L}\,\supset\,  - Y_\nu N \left(\tilde{\epsilon} H^{\ast}\right)^{\dag} L
- Y_N S N N + \textrm{H.c.}\,,
\ee
where we have used two component spinor notation and $\tilde{\epsilon} = i \sigma_2$. 
Spinor and SU(2) indices are understood 
to be contracted trivially, following matrix multiplication rules.
$H$ and $L$ are  
the SM Higgs and lepton doublets, respectively, and $N$ is the SM-singlet right-handed neutrino. 
$Y_\nu$ and $Y_N$ are typically 3-by-3 matrices in flavor space. 
Here, for simplicity, we will only focus on a single entry of each matrix: 
$y_{\nu}\equiv (Y_\nu)_{33}$ and $y_{N}\equiv (Y_N)_{33}$\,. 
When the scalar $S$ acquires a vev, the right-handed neutrino develops a Majorana mass term. 

The abelian gauge part of the Lagrangian takes the form
\bea\label{eq:lagE}
\mathcal{L}&\supset& -\frac{1}{4}B_{\mu\nu}B^{\mu\nu}-\frac{1}{4}X_{\mu\nu}X^{\mu\nu}-\frac{\epsilon}{2} B_{\mu\nu}X^{\mu\nu}\nonumber\\
& &\qquad  +i\bar{f}\left(\partial^\mu-i g_Y Q_Y \tilde{B}^\mu-i g_{B-L} Q_{B-L} \tilde{X}^\mu\right)\gamma_\mu f\,,
\eea
where we indicate with 
$\tilde{B}^\mu$ and $\tilde{X}^\mu$ the gauge bosons of U(1)$_Y$ and U(1)$_{B-L}$, respectively, 
and $B_{\mu\nu}$ and $X_{\mu\nu}$ are the corresponding field strength tensors. 
SM fermions $f$ transform under both symmetry factors with charges $Q_Y$, $Q_{B-L}$ so that
kinetic mixing $\epsilon$ is generated between the two abelian groups.

It is convenient to work in a basis in which the gauge fields are canonically normalized. This can be achieved by a rotation\cite{Holdom:1985ag,Babu:1996vt}
\be
\begin{pmatrix}
\tilde{B}^\mu\\
\tilde{X}^\mu
\end{pmatrix}
=\begin{pmatrix}
1 && -\epsilon/\sqrt{(1-\epsilon^2)} \\
0 && 1/\sqrt{(1-\epsilon^2)}  
\end{pmatrix}
\begin{pmatrix}
V^\mu\\
D^\mu
\end{pmatrix}\,,
\ee
which parameterizes the gauge interaction vertices of Lagrangian~(\ref{eq:lagE}) in terms of a ``visible'' gauge boson $V^{\mu}$
and a ``dark'' gauge boson $D^{\mu}$:
\be\label{mix:vertex}
(Q_Y\,,Q_{B-L})
\left(\begin{array}{cc}
g_{Y} & 0 \\ 0  & g_{B-L} 
\end{array} \right)
\left(\begin{array}{c}
\tilde{B}^\mu\\ \tilde{X}^\mu
\end{array} \right)\quad \to \quad
(Q_Y\,,Q_{B-L})
\left(\begin{array}{cc}
g_{Y} & g_{\epsilon} \\ 0  & g_{d} 
\end{array} \right)
\left(\begin{array}{c}
V^\mu\\ D^\mu
\end{array} \right)\,.
\ee
The elements $g_Y$, $g_d$, and $g_{\epsilon}$ 
are related to the original couplings as
\be\label{eq:gcoups}
g_Y\to g_Y,\quad g_d=\frac{g_{B-L}}{\sqrt{1-{\epsilon}^2}}\,,\quad g_{\epsilon}=-\frac{\epsilon\, g_Y}{\sqrt{1-\epsilon^2}}\,.
\ee

The value of the gravitational parameter $f_g$ is determined by equating the 1-loop 
RG flow of the hypercharge gauge coupling onto the low-scale $\overline{\text{MS}}$ value in the SM, in agreement with assumption~1 in Sec.~\ref{sec:gen_not}. We take $g_Y^{\text{SM},\overline{\text{MS}}}(M_t=173.1\gev)=0.36$\cite{Buttazzo:2013uya}.  
We extract $f_y$ by equating the 1-loop flow of the top Yukawa coupling to its $\overline{\text{MS}}$ value,
$y_t^{\text{SM},\overline{\text{MS}}}(M_t)=0.95$\cite{Workman:2022ynf}, in agreement with
assumption 2 in Sec.~\ref{sec:gen_not}. We obtain 
\be\label{eq:fgBML}
f_g\left(\textrm{1 loop}\right) = 0.0097\,, \quad f_y\left(\textrm{1 loop}\right) = 0.0020\,,  
\ee
which lead to the following predictions at the irrelevant fixed point:
\begin{equation}\label{eq:gau1_fp}
    g_Y^{\ast} \left(\text{1 loop}\right) = 0.4734\,, \qquad \qquad g_d^{\ast} \left(\text{1 loop}\right)= 0.4420\,, \qquad \qquad g_\epsilon^{\ast} \left(\text{1 loop}\right)= -0.3450\,,
\end{equation}
\begin{equation}\label{eq:yuk1_fp}
    y_t^{\ast} \left(\text{1 loop}\right) = 0.2901\,, \qquad \qquad y_{\nu}^{\ast} \left(\text{1 loop}\right)= 0.5398\,,\qquad \qquad y_N^{\ast} \left(\text{1 loop}\right)=0.3868\,.
\end{equation}

The subsequent low-scale predictions for the dark sector coupling and kinetic mixing, at the scale $M_t$, are
\be\label{eq:gau1_ew}
g_d\left(M_t, \text{1 loop}\right) = 0.3343\,, \qquad \qquad g_{\epsilon}\left(M_t, \text{1 loop}\right) = -0.2609\,,
\ee
\be
y_{\nu}\left(M_t, \text{1 loop}\right) =0.4790\,, \qquad \qquad y_N\left(M_t, \text{1 loop}\right) = 0.3392\,.
\ee
For practicality reasons, we choose to present the low-scale predictions at the EWSB scale as reference. In a realistic neutrino-mass model one should decouple the NP at the scale of the vev of the scalar field $S$. Note, however, that the dimensionful parameters of the Lagrangian are canonically relevant and as such they 
do not emerge as predictions of the fixed-point analysis. 

\begin{figure}[t]
	\centering%
    \subfloat[]{%
		\includegraphics[width=0.41\textwidth]{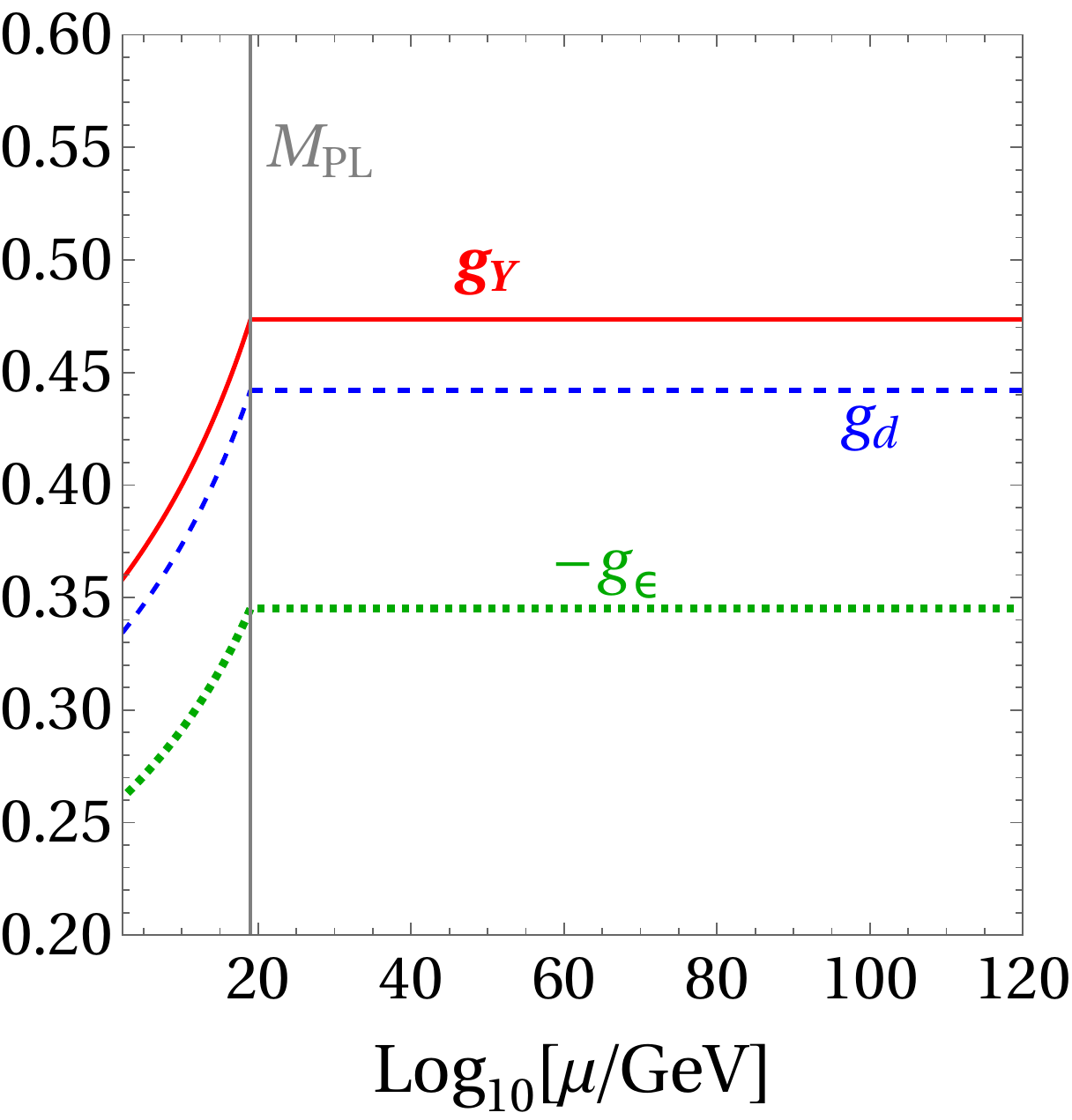}}
    \hspace{1cm}
    \subfloat[]{%
		\includegraphics[width=0.4\textwidth]{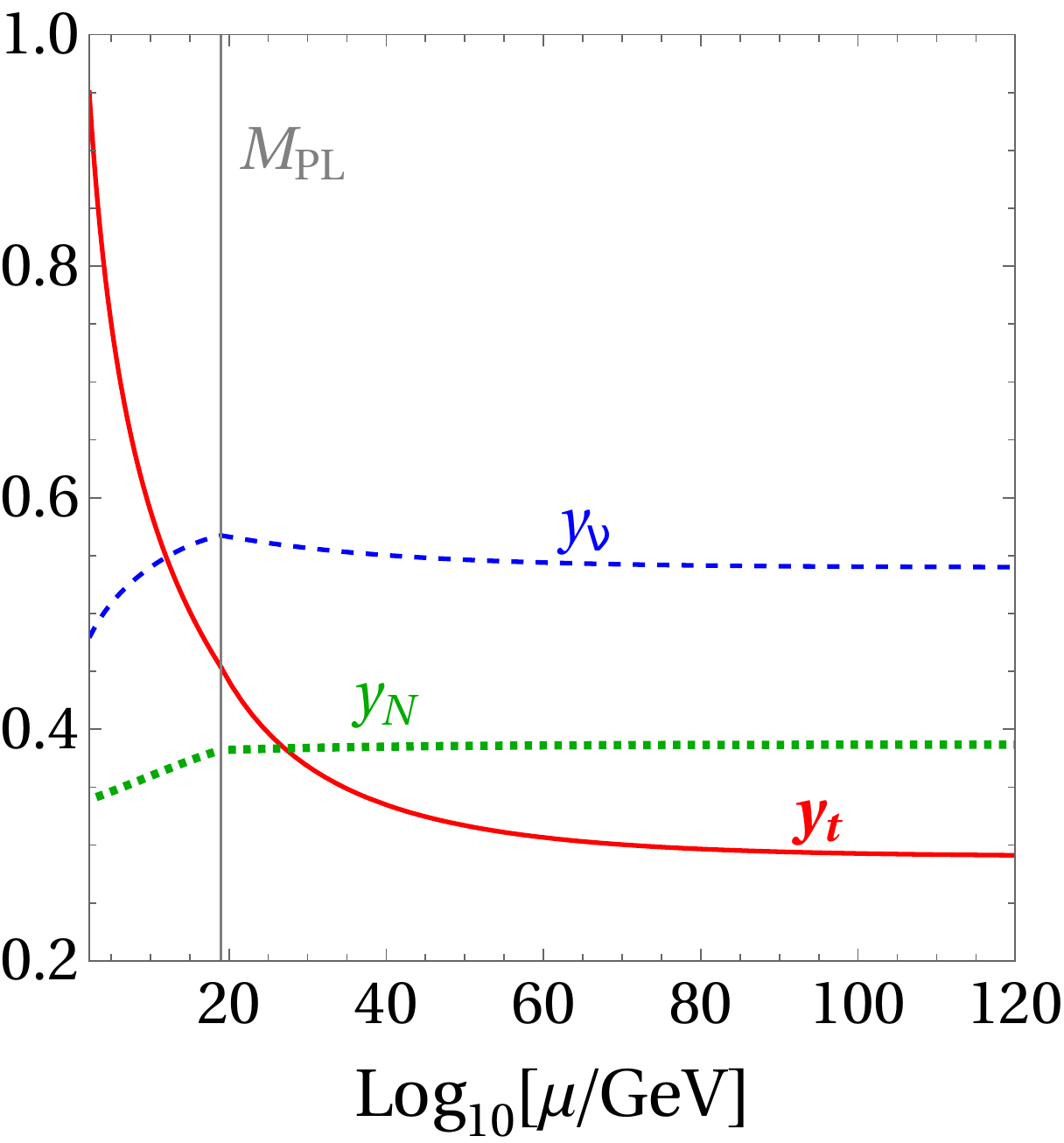}}
\caption{RG flow of the (a) gauge and (b) Yukawa couplings in the gauged $B-L$ model from trans-Planckian energies down to the EWSB scale, for a scenario characterized by the UV fixed points of \refeq{eq:gau1_fp} and \refeq{eq:yuk1_fp}. The gray vertical line indicates the fixed position of the Planck scale, $\mpl=10^{19}\gev$.} 
\label{fig:bl_1loop}
\end{figure}

In \reffig{fig:bl_1loop}(a) we show for illustration the RG flow of the three gauge couplings from their UV fixed point, cf.~\refeq{eq:gau1_fp}, down to the EWSB scale, cf.~\refeq{eq:gau1_ew}. Gravity decouples at the scale $\mpl=10^{19}\gev$, marked as a vertical gray line. The RG flow of the three Yukawa couplings is presented in \reffig{fig:bl_1loop}(b). Note the deviation of $y_t$ from its UV fixed-point value, due to the presence of relevant directions associated with the gauge couplings $g_3$ and $g_2$. 

\subsection{Leptoquark $S_3$\label{sec:LQ}}

The SM is extended in this case by a complex scalar that carries both lepton and baryon number\cite{Buchmuller:1986zs,Davies:1990sc}. 
Its SU(3)$_{\textrm{c}}\times$SU(2)$_L\times$U(1)$_Y$
quantum numbers are
\be
S_3 : \left(\mathbf{\bar{3}},\mathbf{3},1/3\right).
\ee

Following the notation of Ref.\cite{Kowalska:2020gie}, 
we introduce the Yukawa-type interaction of $S_3$ with SM fermions as follows
\begin{align}
    \mathcal{L} \supset - Y_{\textrm{LQ}}\, Q^T \tilde{\epsilon} S_3\, L + \textrm{H.c.}\,,
\end{align}
where, again, spinor and SU(2) indices are contracted trivially, 
$Q$ and $L$ are  
the SM quark and lepton doublets, respectively,
and $Y_{\textrm{LQ}}$ is a 3-by-3 matrix in flavor space. 
For simplicity we shall focus only on the 3rd generation, denoting $y_{\textrm{LQ}}\equiv (Y_{\textrm{LQ}})_{33}$\,.

As before, the gravitational parameters $f_g$ and $f_y$ are determined 
by equating the 1-loop RG flow of the hypercharge gauge coupling and the top Yukawa coupling
onto their low-scale $\overline{\text{MS}}$ values. 
The resulting gravitational parameters read 
\be\label{eq:fgS3}
f_g \left(\text{1 loop}\right)=0.0106\,,\qquad f_y \left(\text{1 loop}\right)=-0.0004\,,
\ee
leading to the following fixed-point values for the irrelevant couplings:
\be\label{eq:lep1_fp}
g_Y^\ast \left(\text{1 loop}\right)=0.4823\,,\qquad y_t^\ast \left(\text{1 loop}\right)=0.2340\,,\qquad y^\ast_{\textrm{LQ}} \left(\text{1 loop}\right)=0.1132\,.
\ee
Finally, the low-scale prediction for the LQ coupling $y_{\textrm{LQ}}$ reads
\be\label{eq:lep1_ew}
y_{\textrm{LQ}}\left(M_t, \text{1 loop}\right)=0.4270\,.
\ee

\section{Estimation of the uncertainties}
\label{sec:uncertainty}

The predictions from AS for a specific NP 
model find perhaps their greatest usefulness when deriving phenomenological bounds on the relevant parameters of the model in question (\textit{e.g.} its masses)\cite{Reichert:2019car,Eichhorn:2020kca,Kowalska:2022ypk} or, alternatively, when  
confronting the model with an experimental anomaly\cite{Kowalska:2020gie,Kowalska:2020zve,Chikkaballi:2022urc}. 
In particle physics phenomenology it is natural to interpret a new experimental bound (or, in alternative, 
a deviation from the SM emerging in the precision measurement of a certain process) as an estimate of the NP contribution to the Wilson coefficient of a higher-dimensional operator in the effective field theory~(EFT). While the size of a Wilson coefficient generally points to a UV scale of interest, this corresponds to the actual mass 
of the particles of the UV completion only if the latter generate  
the Wilson coefficient at the tree level with couplings of order one. 
In practice the picture is more obscure, due to our complete ignorance of the actual type and strength of UV interactions. 

More specifically, the generic EFT matching of a Wilson coefficient $\mathcal{C}_{\textrm{NP}}$ 
to a UV Lagrangian with couplings $c_i$ of unknown strength depends possibly on several factors:
\be\label{eq:genmat}
\frac{\mathcal{C}_{\textrm{NP}}}{\Lambda^n}\approx \frac{c_i c_j}{m_{\textrm{NP}}^n}\times \textrm{loop factor}\times \frac{c_k v_H}{m_{NP}}\times ...\,,
\ee 
where $\Lambda$ sets the scale of the UV physics that is integrated out and 
the last factor, containing the Higgs vev $v_H$, is meant to indicate effects that depend explicitly on the breaking of electroweak symmetry, \textit{e.g.}, the generation of a neutrino mass in see-saw models, or the 
presence of a chiral enhancement in dipole operators. As was discussed in Sec.~\ref{sec:predictions}, 
the assumption of AS can lead to predictions for the  dimensionless parameters of the Lagrangian that are not constrained otherwise. Thus, once the l.h.s.~of \refeq{eq:genmat} is determined following a measurement at the low scale, 
$m_{\textrm{NP}}$ remains the only unknown variable on the r.h.s., which can be extracted straightforwardly. 
Following this procedure, under the three assumptions of Sec.~\ref{sec:gen_not}, in Ref.\cite{Kowalska:2020gie} a fairly precise prediction for the mass of 
the $S_3$ LQ was obtained from a global fit of the Wilson coefficients of the Weak EFT; in Ref.\cite{Kowalska:2020zve} predictions for the mass of two vector-like fermions were obtained from 
the measurement of leptonic dipole operators; and in Ref.\cite{Kowalska:2022ypk} a prediction for the Majorana mass of sterile neutrinos was obtained from the see-saw mechanism. 

In this context, pinning down the theoretical uncertainties that mar the predicted couplings~$c_i$,
which are then fed into \refeq{eq:genmat} to obtain $m_{\textrm{NP}}$, becomes of crucial importance.     
Ideally, the error should remain significantly below the experimental uncertainty on the determination of the Wilson coefficient $\mathcal{C}_{\textrm{NP}}$. To get a quantitative sense of the precision we expect from the asymptotically safe prediction, we point out that observing, for example, 
an hypothetical deviation from the SM at the $p\,\sigma$ level in some experiment, would induce a relative uncertainty on a related Wilson coefficient $\mathcal{C}_{\textrm{NP}}$ around its central value, at the level of 
\be\label{eq:operr}
\frac{\delta \mathcal{C}_{\textrm{NP}}}{\mathcal{C}_{\textrm{NP}}}\approx \frac{1}{p}.
\ee
Thus, in order for this approach to make sense, 
we expect the theoretical uncertainty on the coupling prediction to remain below the few percent level. 

\subsection{Impact of higher-order corrections to the beta functions\label{sec:higher}}

In this subsection, we describe how the predictions we obtained in Secs.~\ref{sec:gBmL} and \ref{sec:LQ} 
are modified when we drop the first approximation listed in Sec.~\ref{sec:intro}. In other words, we consider the DREG beta functions in the matter sector 
at perturbation orders higher than~1.  Note, however, that in order to ensure maximal predictivity assumptions~1, 2, and 3 in Sec.~\ref{sec:gen_not} must hold throughout this analysis.\footnote{In Sec.~\ref{sec:abel_gau} we will briefly describe an additional source of uncertainty arising when one relaxes assumption~1 in Sec.~\ref{sec:gen_not}, \textit{i.e.}, when the system of gauge couplings features only relevant directions.}

\paragraph{Comment on threshold corrections}
Since we impose initial conditions on the solutions of 2-loop RGEs for the SM couplings, consistency requires that we include threshold corrections at the 1-loop order.
Such corrections would also reduce the uncertainty associated with the unphysical matching scale.

In the case of an unbroken gauge symmetry, a generic expression for threshold correction exists. For example, in the $S_3$ LQ model the matching condition for $g_3$ yields (see for example\cite{Athron:2012pw})
\be\label{eq:tc_g3}
g_3^{\text{NP}} = g_3^{\text{SM}} - \frac{g_3^3}{(4\pi)^2} \frac{1}{6} \left( \ln \frac{m_{\phi_{1/3}}}{\mu} + \ln \frac{m_{\phi_{4/3}}}{\mu} + \ln \frac{m_{\phi_{-2/3}}}{\mu}\right).
\ee
Assuming, as we do in this work, that the matching scale is set to $\mu = M_t$, this correction is small even for leptoquarks with masses of a few TeV. If, for example,  $m_{\phi_{1/3}} = m_{\phi_{4/3}} = m_{\phi_{-2/3}} = 2$ TeV, the relative shift due to \refeq{eq:tc_g3} is smaller than 1\%. 
In the case of a broken gauge symmetry the expression of threshold corrections is more involved, but the final result is similar to \refeq{eq:tc_g3} in its analytical form, being a sum of terms involving logs of the NP particle masses. Therefore, as long as the NP scale is not much different from the matching scale, threshold corrections remain small.

Matching conditions for the SM Yukawa couplings are obtained by matching the SM fermion masses, which are measured directly. For example, for the top Yukawa coupling one obtains\cite{Athron:2012pw}
\be
y_t^{\text{NP}} = \frac{g_2^\text{NP} M_t}{\sqrt{2} M_W} \left(1 - \frac{\delta M_W}{M_W} + \frac{\delta M_t}{M_t} \right),
\ee
where $M_W$ is the measured mass of the $W$ boson and $\delta$ denotes the finite part of the full self energy.
To quantify the impact of threshold corrections on $y_t$ we have created an $S_3$ LQ model in \texttt{FlexibleSUSY}\,\cite{Staub:2009bi,Staub:2010jh,Staub:2012pb,Staub:2013tta,Allanach:2001kg,Athron:2014yba,Athron:2017fvs}.
For a parameter point with $m_{S_3}^2 = 1$ TeV, $\lambda_{S_3} = \lambda_{HS_3} = 0.1$ (see Eq.\,\eqref{eq:S3_potential} in Appendix~\ref{app:LQRGE} for the definition of those parameters) and $y_{\text{LQ}} = 1$ we obtain $y_t^\text{NP} = 0.92$, as opposed to 0.95 which is the value we use in this work.

The correct matching of the NP theory to the SM is required to make precise statements about the values of its parameters.
This however depends on the mass spectrum of the model and possibly on mixing matrices and cannot be done without discussing also the dimensionful parameters of the matter theory, which are canonically relevant at the fixed point and therefore cannot be predicted. Since none of the statements made in this work depend on
the treatment of threshold corrections, we do not lose generality by neglecting them. In the numerical analysis we will therefore equate the NP equivalents of the SM parameters to their SM values (as was done in Sec.~\ref{sec:predictions}) also at the 2-loop level.

\paragraph{Gauge couplings} Given assumptions~1 and 3 in Sec.~\ref{sec:gen_not}, one can estimate 
the numerical value of the gravitational correction $f_g$ at the fixed point. Once this is extracted, the property of universality in gravitational interactions can be invoked to find the irrelevant fixed points of other gauge couplings. 
As was discussed above, the non-abelian gauge couplings remain relevant, so that we can limit our analysis to the generic system of abelian couplings. In agreement with the $B-L$ example, let us consider U(1)$_Y\times$U(1)$_{B-L}$. 
It is straightforward to extend our conclusions to any pair of abelian symmetries. As can be seen in Appendix~\ref{app:rges}, the 
pertinent RGEs take the following parametric form, common to all models with abelian mixing: 
\bea
\frac{d g_Y}{dt}&=&\frac{1}{16 \pi^2} \left(b_Y+\Pi_{n\geq 2}^{(Y)} \right) g_Y^3
-f_g\, g_Y\label{eq:gyRGEKM}  \\
\frac{d g_d}{dt}&=&\frac{1}{16 \pi^2}\left[ \left(b_Y+\Pi_{n\geq 2}^{(Y)}\right) g_d g_{\epsilon}^2+  \left(b_d+\Pi_{n\geq 2}^{(d)}\right)  g_d^3+ \left(b_\epsilon+\Pi_{n\geq 2}^{(\epsilon)}\right)  g_d^2g_\epsilon\right]
-f_g\, g_d\label{eq:giRGEKM} \\
\frac{d g_\epsilon}{dt}&=& \frac{1}{16 \pi^2}\left[  \left(b_Y+\Pi_{n\geq 2}^{(Y)}\right) \left( g_\epsilon^3+2 g_Y^2 g_\epsilon  \right) + \left(b_d+\Pi_{n\geq 2}^{(d)}\right) g_d^2 g_\epsilon\right. \nonumber \\
 & & \left.+\left(b_\epsilon+\Pi_{n\geq 2}^{(\epsilon)}\right) \left(g_Y^2 g_d + g_d g^2_\epsilon \right)\right]-f_g\, g_\epsilon\,.\label{eq:geRGEKM}
\eea
Equations \eqref{eq:gyRGEKM}--\eqref{eq:geRGEKM} are given in terms of the one-loop coefficients $b_Y$, $b_d$, and $b_{\epsilon}$, and generic $n$-loop contributions,
\be \label{eq:Ploopn}
\Pi_{n\geq 2}^{(i)}=\frac{1}{16\pi^2} \sum_{l,k} \alpha_{lk}^{(i)}  c_{l}\, c_{k} 
+ \sum_{n>2} \sum_{l_1... l_{n-1}} \frac{1}{(16\pi^2)^{n-1}}
\alpha_{l_1... l_{n-1}}^{(i)}  c_{l_1}^2 ... c_{l_{n-1}}^2 \,,
\ee
expressed in terms of loop coefficients $\alpha_{l_1... l_{n-1}}^{(i)}$, with $i=Y,d,\epsilon$. 
The $c_l$ parameters indicate the gauge and Yukawa couplings of the theory (at perturbation order $n=2$), or the gauge, Yukawa, and (quartic)$^{1/2}$ couplings (at order $n>2$). We require that Eqs.~\eqref{eq:gyRGEKM}--\eqref{eq:geRGEKM} develop a zero above the Planck scale. 

Assuming for the moment 
that the fixed point for the gauge couplings is developed sharply at a $M_{\textrm{Pl}}=10^{19}\gev$, 
one can express the value of $f_g$ in terms of the (known) U(1)$_Y$ trans-Planckian fixed point $g_Y^{\ast}$, with an accuracy that increases at each successive order,
\be
f_g(n\textrm{ loops})\approx \frac{g_Y^{\ast 2}(n\textrm{ loops})}{16 \pi^2}  \left(b_Y+\Pi_{n\geq 2}^{(Y)\ast} \right)\,,
\ee
where the asterisk refers to all couplings being set at their UV fixed-point value.
The predicted ratios of the gauge couplings 
do not depend explicitly on the value of $f_g$. One can define
\bea
r_{g,d}^{\ast}(n\textrm{ loops})&\equiv&\frac{g_d^{\ast}}{g_Y^{\ast}}(n\textrm{ loops})\approx \frac{2 \tilde{b}_Y}{\sqrt{4 \tilde{b}_Y \tilde{b}_d-\tilde{b}_{\epsilon}^2}}\label{eq:gXFP}\,,\\
r_{g,\epsilon}^{\ast}(n\textrm{ loops})&\equiv&\frac{g_{\epsilon}^{\ast}}{g_Y^{\ast}}(n\textrm{ loops})\approx -\frac{\tilde{b}_\epsilon}{\sqrt{4 \tilde{b}_Y \tilde{b}_d-\tilde{b}_{\epsilon}^2}}\,,\label{eq:gEFP}
\eea
where we have adopted a simplified notation,
\be\label{eq:loopcoe}
\tilde{b}_i \equiv b_i+\Pi_{n\geq 2}^{(i)\ast}\,.
\ee

In order to obtain some quantitative estimates, let us retain for simplicity only 
the 2-loop corrections, $\Pi^{(i)}_2$, and quantify the uncertainty on the ratios $r_{g,i(=d,\epsilon)}^{\ast}$ 
by calculating 
\be\label{eq:delrg}
\frac{\delta r_{g,i}^{\ast}}{ r_{g,i}^{\ast}}=\frac{r_{g,i}^{\ast}(\textrm{2 loops})-r_{g,i}^{\ast}(\textrm{1 loop})}{r_{g,i}^{\ast}(\textrm{1 loop})}\,.
\ee

 \begin{figure}[t]
	\centering%
		\includegraphics[width=0.48\textwidth]{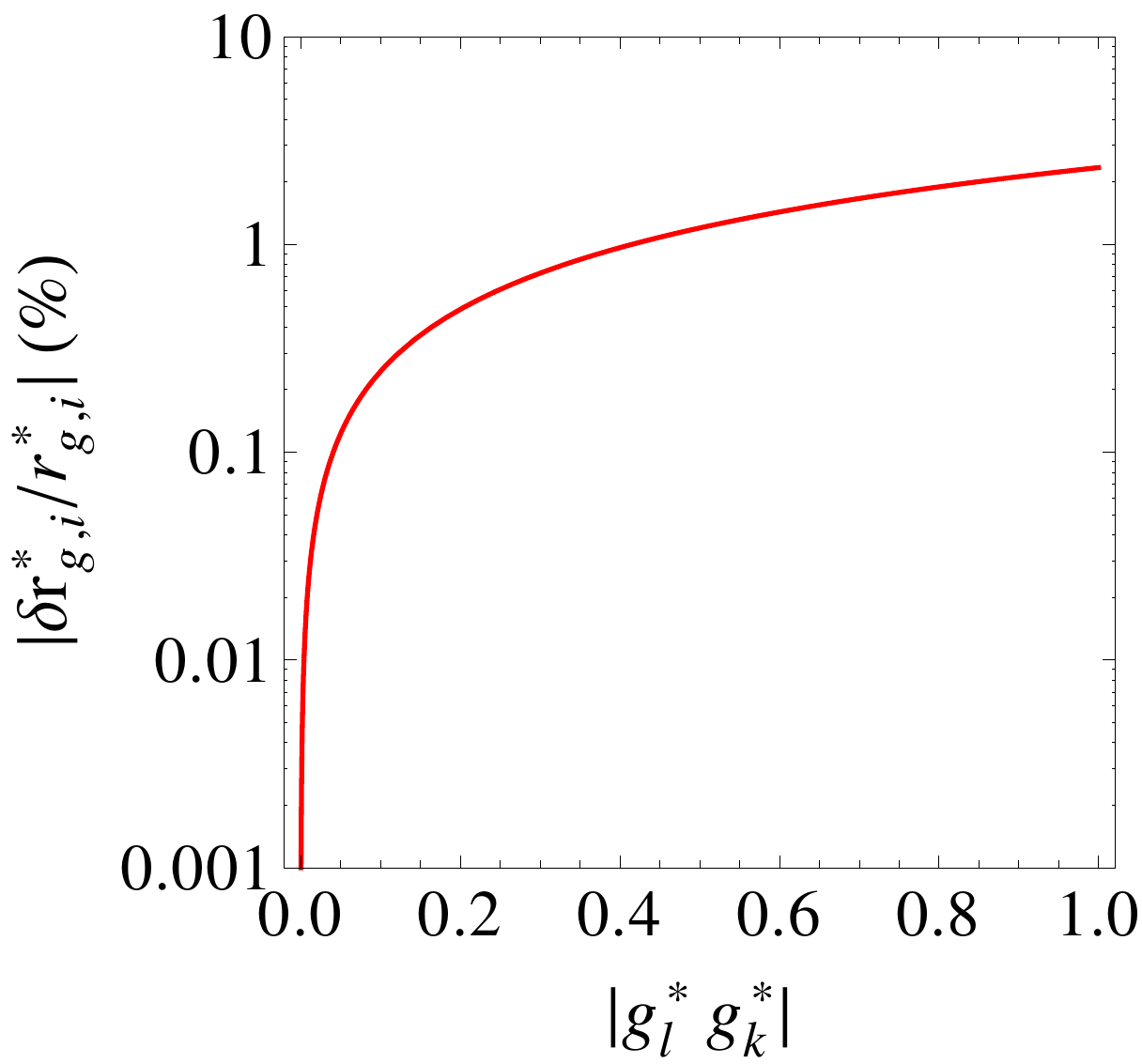}
\caption{
Estimated percent uncertainty on the predicted fixed-point ratios of abelian gauge couplings $|\delta r_{g,i}^{\ast}/ r_{g,i}^{\ast}|$ in the $B-L$ model due to missing higher-loop contributions.
The uncertainty is given as a function of the product of abelian gauge couplings at the fixed point.}
\label{fig:RGEunc}
\end{figure}

In the $B-L$ model, the 1-loop coefficients read $b_Y=41/6$, 
$b_d=12$, $b_\epsilon=32/3$. In general, the dominant coefficients
$\alpha_{lk}^{(i)}$ in \refeq{eq:Ploopn} 
are by far those corresponding to the product of two gauge couplings. Retaining, thus, only the abelian 
gauge-coupling contributions in the 2-loop RGEs of Appendix~\ref{app:rges}, 
and assuming in the first approximation that all gauge couplings acquire 
the same fixed-point value, one can factor out 
the sums of coefficients $\alpha_{lk}^{(i)}$ and compute them separately keeping track of 
the relative coupling signs.
The percent uncertainty is presented in \reffig{fig:RGEunc}.
It is very similar in the two cases $i=d,\epsilon$. It is shown as a function of the generic product of abelian gauge couplings $|g_l^{\ast}\,g_k^{\ast}|$ at the fixed point.  We note that the error remains at the percent level, 
far below the uncertainty typically associated with Wilson coefficients of higher-dimensional operators of the EFT, see \refeq{eq:operr}. Similar levels of uncertainty apply to the UV predictions of gauge couplings in other models with abelian gauge mixing, \textit{e.g.}, those analyzed in Ref.\cite{Chikkaballi:2022urc}. 

The precise numerical calculation of the 2-loop uncertainties for the $B-L$ model is reported in \reftable{tab:predHL}. One obtains $\delta r^\ast_{g,d}/r^\ast_{g,d}=-0.41\%$ and $\delta r^\ast_{g,\epsilon}/r^\ast_{g,\epsilon}=-0.44\%$.

\paragraph{Yukawa couplings} The discussion can be repeated with small modifications for the Yukawa couplings. 
Let us parameterize the full set of Yukawa-coupling RGEs of a generic SM+NP theory in the following way:
\be\label{eq:yrRGE}
\frac{d y_r}{dt}=\frac{y_r}{16\pi^2} \left(\sum_j a_j^{(r)} y_j^2- \sum_{l,k} a_{lk}^{\prime (r)} g_l g_k+\sum_{n\geq 2}\widetilde{\Pi}_n^{(r)}  \right)+\sum_{m,p,q\neq r}\frac{y_m y_p y_q}{16 \pi^2}\left(a^{\prime \prime}_{m p q} +\sum_{n\geq 2} \tilde{\Delta}^{(m p q)}_n\right) -f_y y_r,
\ee
where $y_{r(j)=1,2,\dots}$ label the set of Yukawa couplings in the Lagrangian and $g_{l(k)}$ label the gauge couplings 
of the theory. In \refeq{eq:yrRGE} we have defined the generic \textit{multiplicative} $n$-loop contribution as
\be\label{eq:Piloopn}
\widetilde{\Pi}_n^{(r)} \equiv \frac{1}{(16 \pi^2)^{n-1}} \sum_{l_1 l_2 ... l_{2 n}} \alpha_{l_1 l_2 ... l_{2n}}^{\prime (r)}  c_{l_1} c_{l_2} ... c_{l_{2n}} \,,
\ee
and the generic \textit{additive} $n$-loop piece as
\be\label{eq:Deloopn}
\tilde{\Delta}^{(m p q)}_n \equiv \frac{1}{(16 \pi^2)^{n-1}} 
\sum_{l_1... l_{2 n-n}} \alpha^{\prime \prime (m p q)}_{l_1... l_{2 n-n}} c_{l_1}...c_{l_{2 n-2}}\,,
\ee
in terms of $n$-loop coefficients $\alpha^{\prime(r)}_{l_1...l_{2n}}$, $\alpha^{\prime \prime(mpq)}_{l_1...l_{2n-2}}$, 
and of all gauge, Yukawa, and (quartic)$^{1/2}$ couplings $c_{l_i}$ 
entering the Yukawa RGEs at order $n\geq2$\,.

Neglecting for the moment the additive part of the beta functions, 
let us assume we solve the system~\eqref{eq:yrRGE} when all the beta functions are equal to zero. 
Repeating the steps that led to the 1-loop results in Sec.~\ref{sec:predictions}, 
we express the unknown gravitational contribution $f_y$ 
in terms of one known ``reference'' fixed-point Yukawa coupling $y_1^{\ast}$, which is in general the top quark's.\footnote{Because of assumption 2 in Sec.~\ref{sec:gen_not}, in order to estimate $f_y$ all one needs is a fixed point with an irrelevant SM Yukawa coupling. Which fixed point  works best for the phenomenology is decided on a case-by-case basis.} 

Let us retain, for simplicity, only the 2-loop correction. One gets 
\begin{multline}\label{eq:fyY}
f_y(\textrm{2 loops})\approx \frac{1}{16 \pi^2} \left[y_1^{\ast 2}(\textrm{2 loops}) \times F_0\left(a_j^{(r)}\right)+G_0\left(\sum_{l,k} a^{\prime (r)}_{lk} g_l^{\ast} g_k^{\ast}(\textrm{2 loops})\,;\,a_{j\neq 1}^{(r)}\right)\right. \\
\left. +H_0\left(\widetilde{\Pi}_2^{(r)\ast}\,;\,a_{j\neq 1}^{(r)} \right) \right]\,,
\end{multline}
where $F_0$, $G_0$, and $H_0$ are rational functions of the 
$a_j^{(r)}$ coefficients, with the latter two
being linear in $\sum_{lk} a^{\prime (r)}_{lk} g^{\ast}_l g^{\ast}_k$ and $\widetilde{\Pi}_2^{(r)\ast}$, respectively. Their 
analytic form is model-dependent and is not needed for the discussion. Note that the reference fixed-point value $y^{\ast}_1(\textrm{2 loops})$ is obtained by equating the 2-loop RGE flow to the $\overline{\textrm{MS}}$ values of the couplings at the EWSB scale. As such, it can be significantly shifted with respect to $y^{\ast}_1(\textrm{1 loop})$, mostly due to the contributions of relevant parameters like $g_3$ and $g_2$ to the 2-loops RGEs. 

Let us thus define the shift at the fixed point due to relevant parameters in the flow,
\be\label{eq:2lshift}
y_1^{\ast 2}(\textrm{2 loops})=y_1^{\ast 2}(\textrm{1 loop})+\delta y_1^{\ast 2}\,,
\ee
and derive the parametric expression for the fixed point of a second Yukawa coupling of the irrelevant type. This is  our prediction, which we indicate as $y_2^{\ast}$\,. Its value must be a function of $f_y$. By using \refeq{eq:fyY} we trade
$f_y$ for $y_1^{\ast}$ and plug into the fixed-point solution for $y_2^{\ast}$:
\begin{multline}\label{eq:yukrat}
y_2^{\ast}(\textrm{2 loops})\approx \left[F_1\left(a_j^{(r)}\right) \left(y_1^{\ast 2}(\textrm{1 loop})+\delta y_1^{\ast 2}\right)+G_1\left(\sum_{l,k} a^{\prime (r)}_{lk} g_l^{\ast} g_k^{\ast}\,;\,a_{j\neq 1}^{(r)}\right)\right. \\
\left. + H_1\left(\widetilde{\Pi}_2^{(r)\ast}\,;\,a_{j\neq 1}^{(r)} \right) \right]^{1/2},
\end{multline}
where the $F_1$, $G_1$, are some other rational functions of coefficients of order one, and the $H_1$ function 
carries the 2-loop terms. Using \refeq{eq:Piloopn}, one can see that the 
higher-loop correction~$H_1$ only depends, 
up to order-one coefficients, on sums of fixed-point products $c_{l_1}^{\ast}c_{l_2}^{\ast}c_{l_3}^{\ast}c_{l_4}^{\ast}$. It is thus very similar in size for all the predicted Yukawa couplings -- $y_{2}^{\ast}$,  $y_{3}^{\ast}$, etc. The shift uncertainty $\delta y_1^{\ast 2}$ also enters in the same way in all predicted Yukawa couplings. However, the relative uncertainties affecting the predictions -- $\delta y_2^{\ast}/y_2^{\ast}$, $\delta y_3^{\ast}/y_3^{\ast}$, etc. -- differ for the different Yukawa couplings because they are very sensitive to the actual size of the coupling itself. If one, for example, had obtained at 1 loop $y_{2}^{\ast}\ll y_1^{\ast}, g_k^{\ast}$, that would have implied a precise cancellation (fine tuning) between $F_1(a_j^{(r)})\,y_1^{\ast 2}$ and the $G_1$ function. As a consequence, the uncertainty due to $\delta y_1^{\ast 2}$ and $H_1$ 
on the final result would grow. 
Conversely, obtaining $y_{2}^{\ast}\approx y_1^{\ast}, g_k^{\ast}$ at 1 loop would imply that such prediction can be trusted to a very good approximation.

For a quantitative example, let us choose a fixed point 
with one irrelevant gauge coupling ($l=k=1$) and two Yukawa couplings ($r,j=1,2$). Equation~\eqref{eq:yukrat} becomes
\be\label{eq:yukrat2}
y_2^{\ast}(\textrm{2 loops})\approx \left[\frac{a_1^{(2)}-a_1^{(1)}}{a_2^{(1)}-a_2^{(2)}}\, 
\left(y_1^{\ast 2}(\textrm{1 loop})+\delta y_1^{\ast 2}\right)
+\frac{\left(a_{11}^{\prime (1)}-a_{11}^{\prime (2)}\right) g_1^{\ast 2}+\widetilde{\Pi}_{2}^{(2)\ast}-\widetilde{\Pi}_{2}^{(1)\ast}}{a_2^{(1)}-a_2^{(2)}}\right]^{1/2}.
\ee
This is the case, for example, of the $S_3$ LQ. 
With respect to the notation of Sec. \ref{sec:LQ}, here 
$g_1\equiv g_Y$, $y_1\equiv y_t$, and $y_2\equiv y_{\textrm{LQ}}$\,. All other couplings are chosen relevant and zero at the fixed point. The 1-loop coefficients in this model can be found in Appendix \ref{app:rges} and read $a_1^{(1)}=9/2$, $a_2^{(1)}=3/2$, $a_{11}^{\prime (1)}=17/12$, $a_1^{(2)}=1/2$, $a_2^{(2)}=8$, $a_{11}^{\prime (2)}=5/6$. 

The shift in the top Yukawa fixed point can be computed numerically: $\delta y_1^{\ast 2}\approx -9\times 10^{-3}$. 
This should be summed to the explicit 2-loop contribution to the beta functions, 
$\widetilde{\Pi}_{2}^{(2)\ast}-\widetilde{\Pi}_{2}^{(1)\ast}$, which can be obtained from Appendix \ref{app:rges}. The latter turns out to be approximately one order of magnitude smaller than $\delta y_1^{\ast 2}$ when the quartic couplings of the scalar potential are neglected. If the quartic couplings are assumed to be of order one, we get instead $\widetilde{\Pi}_{2}^{(2)\ast}-\widetilde{\Pi}_{2}^{(1)\ast} \approx -3\times 10^{-2}$. 
The first of these determinations leads to the largest uncertainty on the prediction.  By plugging the 1-loop coefficients  into \refeq{eq:yukrat2}, together with the 1-loop fixed-point values given in Sec. \ref{sec:LQ}, one can easily compute the percent deviation,
\be\label{eq:deltay}
\frac{\delta y_2^{\ast}}{y_2^{\ast}}=\frac{y_2^{\ast}(\textrm{2 loops})-y_2^{\ast}(\textrm{1 loop})}{y_2^{\ast}(\textrm{1 loop})}\approx -25\%\,.
\ee
We note that in this example the uncertainty is not negligible. 
This was expected, since at 1 loop one finds 
$y_{2}^{\ast}< y_1^{\ast}, g_1^{\ast}$, see the values in Sec.~\ref{sec:LQ}. 
In the $B-L$ model, on the other hand, one finds $y_{2}^{\ast}, y_{3}^{\ast} \approx 
y_1^{\ast}, g_{k=1,2,3}^{\ast}$, and the uncertainty is smaller, cf.~\reftable{tab:predHL}.

The discussion can be extended, finally, to cases in which  
an additive contribution to the Yukawa coupling RGEs -- the fourth and fifth addends in \refeq{eq:yrRGE} -- 
is present. It is not difficult to realize 
that in those cases the Yukawa 
system only admits a set of real interactive solutions $y_r^{\ast}\neq 0$
of the same order of magnitude as the irrelevant gauge-coupling fixed point. Since there is no fine tuning involved, such solutions are not very sensitive to the $n$-loop corrections.

\setlength\tabcolsep{0.20cm}
\begin{table}[t]\footnotesize
\begin{center}
\begin{tabular}{|c|c|c|cc|ccc|cc|}
\hline
\multirow{4}{*}{$\boldsymbol{B-L}$} & $f_g$ & $g_Y^\ast$ & $g_d^\ast$ & $g_\epsilon^\ast$ & $\delta g_Y^\ast/g_Y^\ast$ & $\delta g_d^\ast/g_d^\ast$ & $\delta g_\epsilon^\ast/g_\epsilon^\ast$ & $\delta g_d/g_d(M_t)$ & $\delta g_\epsilon/g_\epsilon(M_t)$ \\
\cline{2-10}
& 0.0098 & 0.4748 & 0.4415 & $-0.3445$ & $0.3\%$ & $-0.1\%$ & $-0.1\%$ & $-0.4\%$ & $-0.5\%$ \\
\cline{2-10}
 & $f_y$ & $y_t^\ast$ & $y_\nu^\ast$ & $y_N^\ast$ & $\delta y_t^\ast/y_t^\ast$ & $\delta y_\nu^\ast/y_\nu^\ast$ & $\delta y_N^\ast/y_N^\ast$ & $\delta y_\nu/y_\nu(M_t)$ & $\delta y_N/y_N(M_t)$ \\
\cline{2-10}
& 0.0016 & 0.2727 & 0.5220 & 0.3813 & $-6.0\%$ & $-3.3\%$ & $-1.4\%$ & $-1.4\%$ & $-0.8\%$ \\
\hline\hline
\multirow{2}{*}{$\boldsymbol{S_3}$ \textbf{LQ} } & $f_y$ & $y_t^\ast$ & \multicolumn{2}{c|}{$y_{\textrm{LQ}}^\ast$} & $\delta y_t^\ast/y_t^\ast$ & \multicolumn{2}{c|}{  $\delta y_{\textrm{LQ}}^\ast/y_{\textrm{LQ}}^\ast$} & \multicolumn{2}{c|}{  $\delta y_{\textrm{LQ}}/y_{\textrm{LQ}}(M_t)$} \\
\cline{2-10}
 & $-0.0007$ & 0.2133 & \multicolumn{2}{c|}{0.0855} & $-8.8\%$ & \multicolumn{2}{c|}{$-24.5\%$} & \multicolumn{2}{c|}{$-14.3\%$} \\\hline
\end{tabular}
\caption{2-loop determination of the gravity parameters $f_g$ and $f_y$, fixed-point values of the reference and the to-be-predicted couplings, percent uncertainty at the fixed point, and percent uncertainty at the low scale for the models introduced in Sec.~\ref{sec:predictions}. The uncertainties are defined w.r.t.~the 1-loop results of Sec.~\ref{sec:predictions}, cf.~\refeq{eq:deltay}. 
}
\label{tab:predHL}
\end{center}
\end{table}

\paragraph{Numerical examples}

We provide quantitative estimates of the higher-loop effects discussed in this section in the two NP models introduced in Sec.~\ref{sec:predictions}. 

In \reftable{tab:predHL} we present the 2-loop determinations of the gravity parameters $f_g$ and $f_y$, the fixed-point values of the model couplings, as well as their percent uncertainties at the fixed point and at the low-energy scale, $\mu=M_t$. The latter are derived under the assumption that the low-scale value of the reference SM coupling ($g_Y(M_t)$ in the $B-L$ model, $y_t(M_t)$ in the Yukawa sector of the $B-L$ model 
and in the $S_3$ LQ model) is not varied when performing the fixed point analysis at 2 loops. 

The results shown in \reftable{tab:predHL} confirm our discussion. The impact of higher-order corrections on the fixed-point value of a coupling is almost negligible for the gauge couplings, while in the Yukawa sector it depends on the relative size of the reference and the predicted couplings. In the $B-L$ model, the fixed-point values of all three couplings are comparable in size. Hence, 
the impact of 2-loop corrections is small and the uncertainty on the NP couplings is of the order of a few percent. Conversely, in the $S_3$ LQ model, $y_{\textrm{LQ}}$ is smaller by a factor of $2$ than $y_t$. The impact of the 2-loop corrections is then more pronounced, and the resulting fixed-point uncertainty increases.

 \begin{figure}[t]
	\centering%
    \subfloat[]{%
		\includegraphics[width=0.4\textwidth]{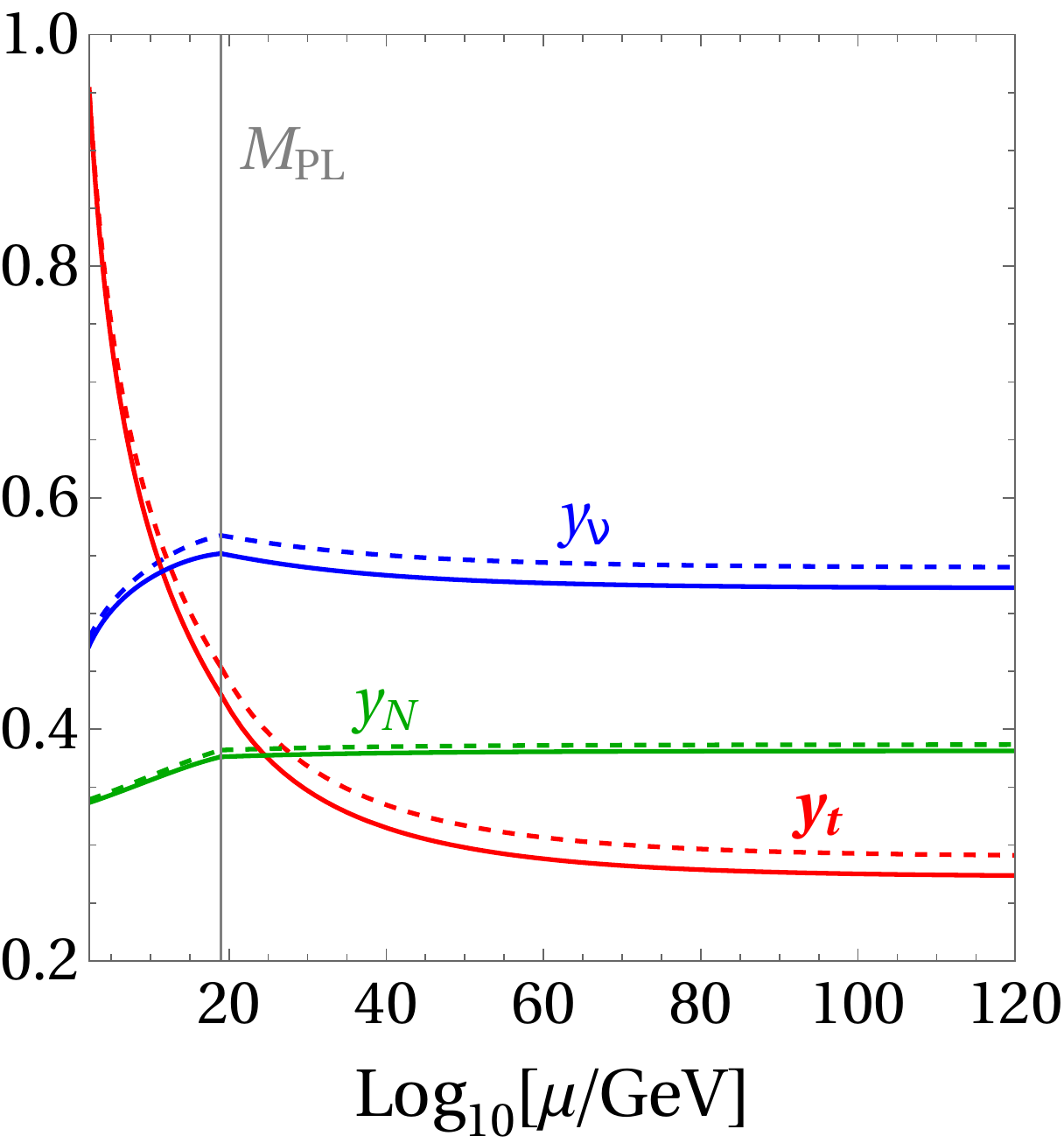}}
      \hspace{1cm}
    \subfloat[]{%
		\includegraphics[width=0.4\textwidth]{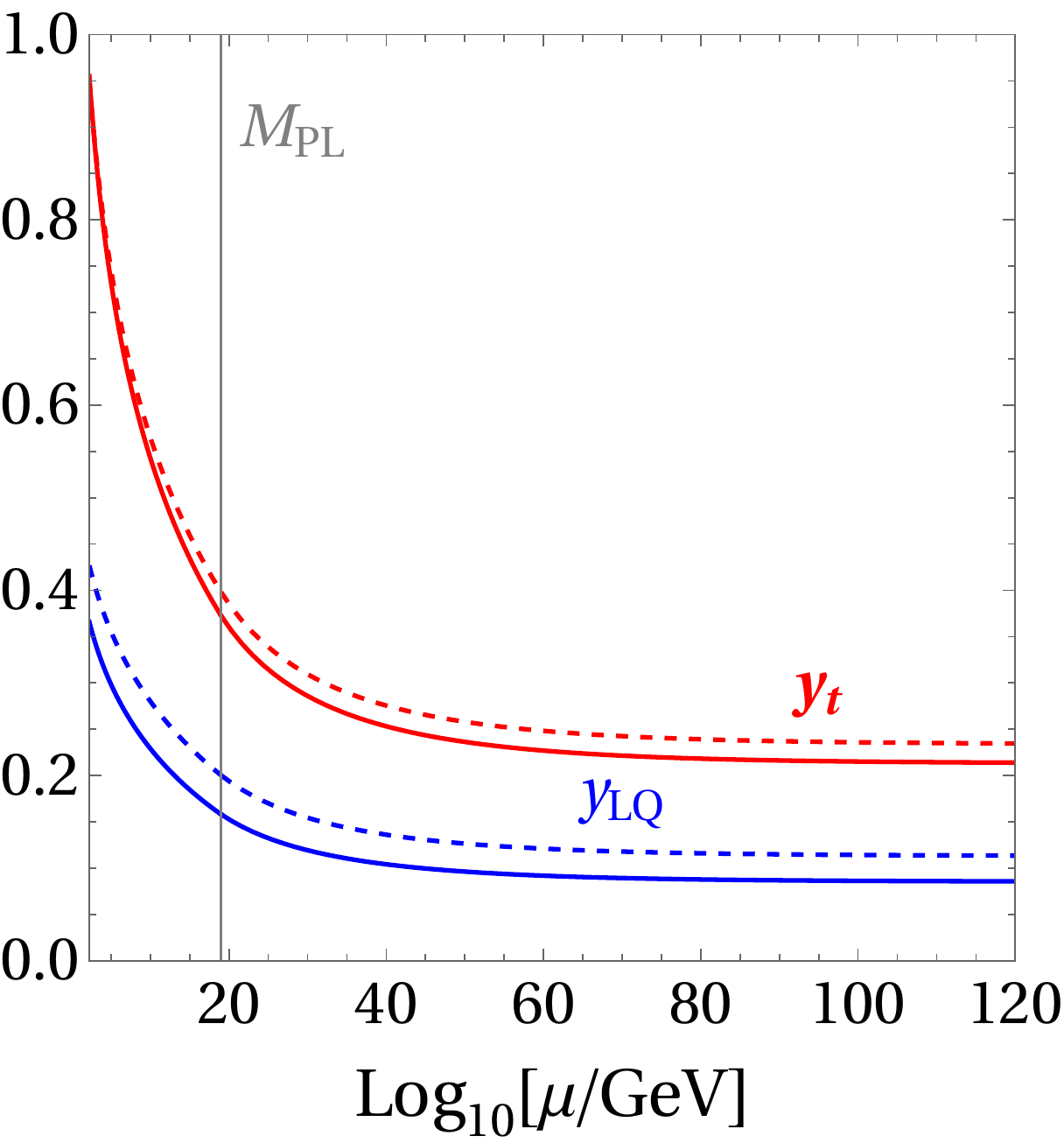}}
\caption{(a) RG flow of the top (red), Dirac neutrino (blue) and sterile neutrino (green) Yukawa couplings of the gauged $B-L$ model at the 1-loop (dashed) and 2-loop (solid) orders. The gray vertical line indicates the fixed position of the Planck scale, $\mpl=10^{19}\gev$. (b) Same as in (a), but for the top (red) and LQ (blue) Yukawa couplings of model $S_3$.}
\label{fig:unc_2loop}
\end{figure}

Closer inspection of the last column of \reftable{tab:predHL} reveals that the low-scale uncertainties of the NP Yukawa couplings are smaller than those at the fixed-point. This is not a coincidence and it is intrinsically related to modifications of the RG flow due to higher-order corrections. To get a better understanding of the underlying mechanism, we show in \reffig{fig:unc_2loop}(a) the 1-loop (dashed) and 2-loop (solid) running of $y_t$ (red), $y_\nu$ (blue) and $y_N$ (green) in the $B-L$ model. The analogous flow of the top (red) and LQ (blue) Yukawa couplings of the $S_3$ model is presented in \reffig{fig:unc_2loop}(b). Due to the negative contribution of the gauge coupling $g_3$ to the beta functions of the Yukawa couplings of particles carrying the color charge, the running value of $y_t$ at 2~loops is generically smaller than at 1~loop. As was discussed below \refeq{eq:yukrat}, it follows that the fixed-point values of the NP Yukawa couplings also tend to be shifted downwards. The effect is stronger for the couplings that depend on $y_t$ already at 1~loop ($y_\nu$ and $y_{\textrm{LQ}}$). 

On the other hand, since the same reference low-scale value $y_t(M_t)$ is used in the fixed-point analysis at any loop order, the RG trajectories of the NP Yukawa couplings have the tendency to focus towards their 1-loop value in the infrared, which results in a reduction of the uncertainty with respect to the prediction at the fixed point. A generic rule of thumb thus applies: the uncertainty in the determination of the fixed-point value of a NP Yukawa coupling provides an upper bound on the uncertainty of the same prediction at the low scale.

\subsection{Dependence on the position of the Planck scale}\label{sec:planck}

In this subsection, we drop the second of the approximations listed in Sec.~\ref{sec:intro}, \textit{i.e.}, we consider what happens if gravity decouples from the matter RGEs sharply at a scale that differs from $10^{19}\gev$ by a few orders of magnitude.

\paragraph{Gauge couplings} When assessing the impact of the Planck-scale position on the predictions for gauge couplings one should note that an uncertainty on the position of the Planck scale is effectively equivalent to an uncertainty on the fixed-point value of the hypercharge gauge coupling $g_Y^\ast$, hence on $f_g$. On the other hand, since the $f_g$ dependence cancels out from Eqs.~\eqref{eq:gXFP} and \eqref{eq:gEFP}, moving the Planck scale back and forth does not affect the predicted ratios $r_{g,i}^{\ast}$ at the 1-loop order.
This feature is not preserved at higher orders in the perturbative expansion. However, the impact of the Planck scale position remains negligibly small, at about~$0.01\%$.   

\paragraph{Yukawa couplings} On the other hand, since \refeq{eq:yukrat} depends explicitly on the fixed-point values of the couplings, changing the position of the Planck scale will alter the prediction for the Yukawa couplings, as it changes the fixed-point values of the abelian gauge couplings, which we indicate collectively with $g_{k=1,2,...}^{\ast}$, and of the reference Yukawa coupling $y_1^{\ast}$. 

Let us define, with slight abuse of notation, $r_{g,k}^{\ast}=g_k^{\ast}/y_1^{\ast}$, $r_{y,2}^{\ast}=y_2^{\ast}/y_1^{\ast}$,
and the percent uncertainty 
\be\label{eq:delrgMPL}
\frac{\delta r_{g(y),i}^{\ast}}{ r_{g(y),i}^{\ast}}=\frac{r_{g(y),i}^{\ast}(\mpl\neq 10^{19}\gev)-r_{g(y),i}^{\ast}(\mpl=10^{19}\gev)}{r_{g(y),i}^{\ast}(\mpl=10^{19}\gev)}\,.
\ee
Neglecting for this discussion the 2-loop contribution, the percent uncertainty on the predicted Yukawa-coupling ratios propagates in \refeq{eq:yukrat} as
\be\label{eq:r2unc}
\frac{\delta r_{y,2}^{\ast}}{r_{y,2}^{\ast}}=\frac{1}{r_{y,2}^{\ast 2}}\, G_1\left(\sum_{l,k} a^{\prime (r)}_{lk}\, r_{g,l}^{\ast} r_{g,k}^{\ast}\cdot \frac{1}{2}\left[\frac{\delta r_{g,l}^{\ast}}{r_{g,l}^{\ast}}+\frac{\delta r_{g,k}^{\ast}}{r_{g,k}^{\ast}}\right] ;\,a_{j\neq 1}^{(r)}\right)\,.
\ee

It is straightforward to re-frame the two cases described in Sec.~\ref{sec:higher} in light of \refeq{eq:r2unc}. If the predicted Yukawa fixed point is of comparable size to the irrelevant gauge couplings, $r_{y,2}^{\ast}\approx  r_{g,l(k)}^{\ast}$, the uncertainty is dominated by the shifting of the gauge coupling fixed point:
\be\label{eq:delta2_appr}
\left|\frac{\delta r_{y,2}^{\ast}}{r_{y,2}^{\ast}}\right|\approx \left|\frac{\delta r_{g,l(k)}^{\ast}}{r_{g,l(k)}^{\ast}}\right|\,.
\ee
Conversely, if the predicted coupling is small, $r_{y,2}^{\ast}\ll  r_{g,l(k)}^{\ast}$, the relative uncertainty can grow substantially and we lose control of the prediction,
\be\label{eq:delta2_hier}
\left|\frac{\delta r_{y,2}^{\ast}}{r_{y,2}^{\ast}}\right|\gg \left|\frac{\delta r_{g,l(k)}^{\ast}}{r_{g,l(k)}^{\ast}}\right|\,.
\ee

As a quantitative example, let us consider again the simple case of two Yukawa couplings, $y_1^{\ast}$, $y_2^{\ast}$, and one gauge coupling $g_1^{\ast}$ (hypercharge), 
all irrelevant at the fixed point.
Equation~\eqref{eq:r2unc} becomes in this case
\be\label{eq:r2num}
\frac{\delta r_{y,2}^{\ast}}{r_{y,2}^{\ast}}=\frac{r_{g,1}^{\ast 2}}{r_{y,2}^{\ast 2}}\cdot \frac{a_{11}^{\prime (1)}-a_{11}^{\prime (2)}}{a_2^{(1)}-a_2^{(2)}}\cdot \frac{\delta r_{g,1}^{\ast}}{r_{g,1}^{\ast}}\,.
\ee
This is the case, \textit{e.g.}, of the $S_3$ LQ model, whose
1-loop coefficients were given below \refeq{eq:yukrat2}.
One can show numerically (1-loop RGEs will suffice) that an eventual shift in the value of the Planck scale will induce a variation in the fixed point values $g_1^{\ast}$, $y_1^{\ast}$, and, consequently, $r_{g,1}^{\ast}\equiv g_1^{\ast}/y_1^{\ast}$.  For example, let us move the Planck scale to $M_{\textrm{Pl}}=10^{16}\gev$ in one case, and to $M_{\textrm{Pl}}=10^{20}\gev$ in another. 
This results in the changing of $g_1^{\ast}=g_Y^{\ast}$ 
from the value in \refeq{eq:lep1_fp} to 0.45 and 0.49, respectively. One finds
$\delta r_{g,1}^\ast/r_{g,1}^\ast(10^{16})\approx -10\%$ and 
$\delta r_{g,1}^\ast/r_{g,1}^\ast(10^{20})\approx 4\%$\,. We can thus use 
\refeq{eq:r2num} to predict $\delta r_{y,2}^\ast/r_{y,2}^{\ast}\approx -1.63\times\delta r_{g,1}^\ast/r_{g,1}^\ast$, which agrees well with numerical derivations showing 
$\delta r_{y,2}^\ast/r_{y,2}^\ast(10^{16})\approx 18.1\%$ and 
$\delta r_{y,2}^\ast/r_{y,2}^\ast(10^{20})\approx -7.8\%$\,.

\paragraph{Numerical examples}

\setlength\tabcolsep{0.20cm}
\begin{table}[t]\footnotesize
\begin{center}
\begin{tabular}{|c|c|c|cc|ccc|cc|}
\hline
$\boldsymbol{B-L}$ & $f_g$ & $g_Y^\ast$ & $g_d^\ast$ & $g_\epsilon^\ast$ & $\delta g_Y^\ast/g_Y^\ast$ & $\delta g_d^\ast/g_d^\ast$ & $\delta g_\epsilon^\ast/ g_\epsilon^\ast$ & $\delta g_d/g_d(M_t)$ & $\delta g_\epsilon/ g_\epsilon(M_t)$ \\
\hline
$10^{20}$ GeV & 0.0102 & 0.4843 & 0.4522 & $-0.3530$ & $2.3\%$ & $2.3\%$ & $2.3\%$ & $0.0\%$ & $0.0\%$ \\
$10^{16}$ GeV & 0.0086 & 0.4445 & 0.4151 & $-0.3240$ & $-6.1\%$ & $-6.1\%$ & $-6.1\%$ &  $0.0\%$ & $0.0\%$ \\
\hline\hline
 & $f_y$ & $y_t^\ast$ & $y_\nu^\ast$ & $y_N^\ast$ & $\delta y_t^\ast/y_t^\ast$ & $\delta y_\nu^\ast/y_\nu^\ast$ & $\delta y_N^\ast/ y_N^\ast$ & $\delta y_\nu/y_\nu(M_t)$ & $\delta y_N/y_N(M_t)$ \\
\hline
$10^{20}$ GeV & 0.0020 & 0.2914 & 0.5523 & 0.3927 & $0.4\%$ & $2.3\%$ & $1.5\%$ & $1.3\%$ & $0.3\%$ \\
$10^{16}$ GeV & 0.0020 & 0.2869 & 0.5069 & 0.3715 & $-1.1\%$ & $-6.1\%$ & $-4.0\%$ & $-3.7\%$ & $-0.9\%$ \\
\hline\hline
$\boldsymbol{S_3}$ \textbf{LQ}   & $f_y$ & $y_t^\ast$ & \multicolumn{2}{c|}{$y_{\textrm{LQ}}^\ast$} & $\delta y_t^\ast/y_t^\ast$ & \multicolumn{2}{c|}{$\delta y_{\textrm{LQ}}^\ast/y_{\textrm{LQ}}^\ast$} & \multicolumn{2}{c|}{$\delta y_{\textrm{LQ}}/y_{\textrm{LQ}}(M_t)$} \\
\hline
$10^{20}$ GeV & $-0.0006$ & 0.2309 & \multicolumn{2}{c|}{0.1043} & $-1.3\%$ & \multicolumn{2}{c|}{$-7.8\%$} & \multicolumn{2}{c|}{$-5.1\%$}\\
$10^{16}$ GeV & 0.00002 & 0.2422 & \multicolumn{2}{c|}{0.1337} & $3.5\%$ & \multicolumn{2}{c|}{$18.1\%$} & \multicolumn{2}{c|}{$10.1\%$}\\
\hline
\end{tabular}
\caption{Impact of the position of the Planck scale on the determination of the gravity parameters $f_g$ and $f_y$\,, fixed-point values of the
reference and the to-be-predicted couplings, percent uncertainty at the fixed
point, and percent uncertainty at the low scale for the models introduced in Sec.~\ref{sec:predictions}. The uncertainties are defined w.r.t.~the 1-loop results of Sec.~\ref{sec:predictions}, cf.~\refeq{eq:delrgMPL}. } 
\label{tab:predMP}
\end{center}
\end{table}

In \reftable{tab:predMP} we summarize the findings of this subsection for our benchmark models. For two different values of the Planck scale, $\mpl=10^{20}\gev$ and $\mpl=10^{16}\gev$, we show determinations of the gravity parameters $f_g$ and $f_y$, the fixed-point values of the model couplings, their percent uncertainties at the fixed point and the uncertainties at the scale $\mu=M_t$. One can immediately see that all the fixed-point values of the gauge couplings are rescaled by the same amount, so the low-scale predictions are not altered. 

\begin{figure}[p]
	\centering%
     \subfloat[]{%
		\includegraphics[width=0.41\textwidth]{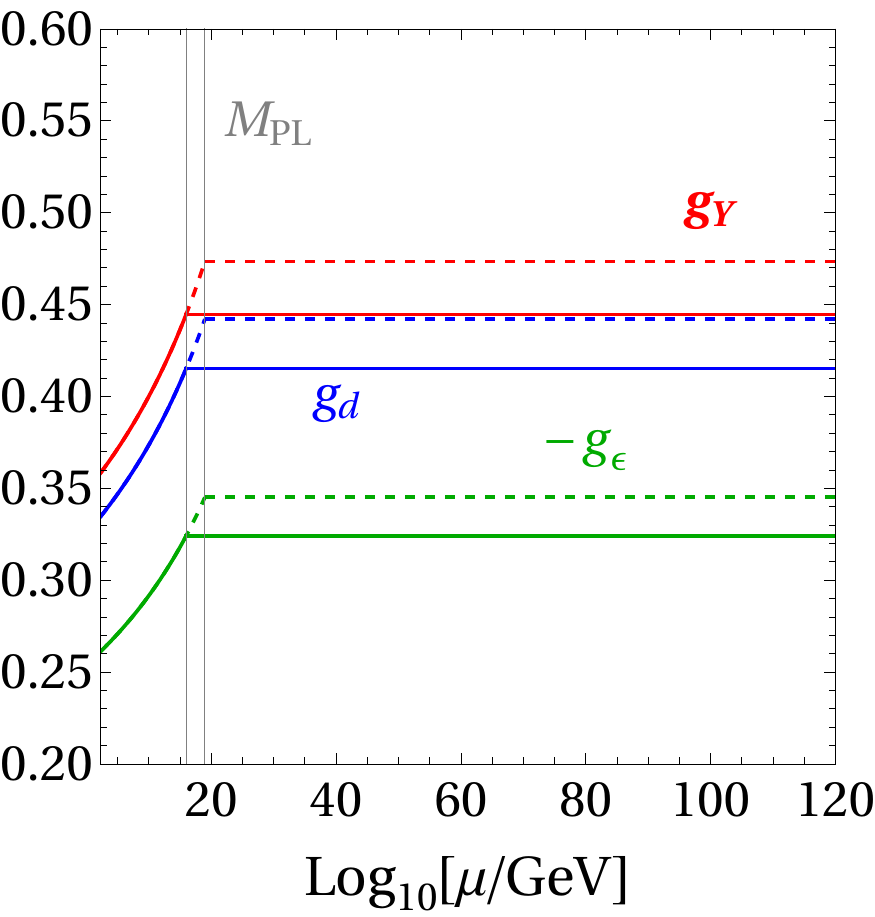}}
 \hspace{1cm}
    \subfloat[]{%
		\includegraphics[width=0.4\textwidth]{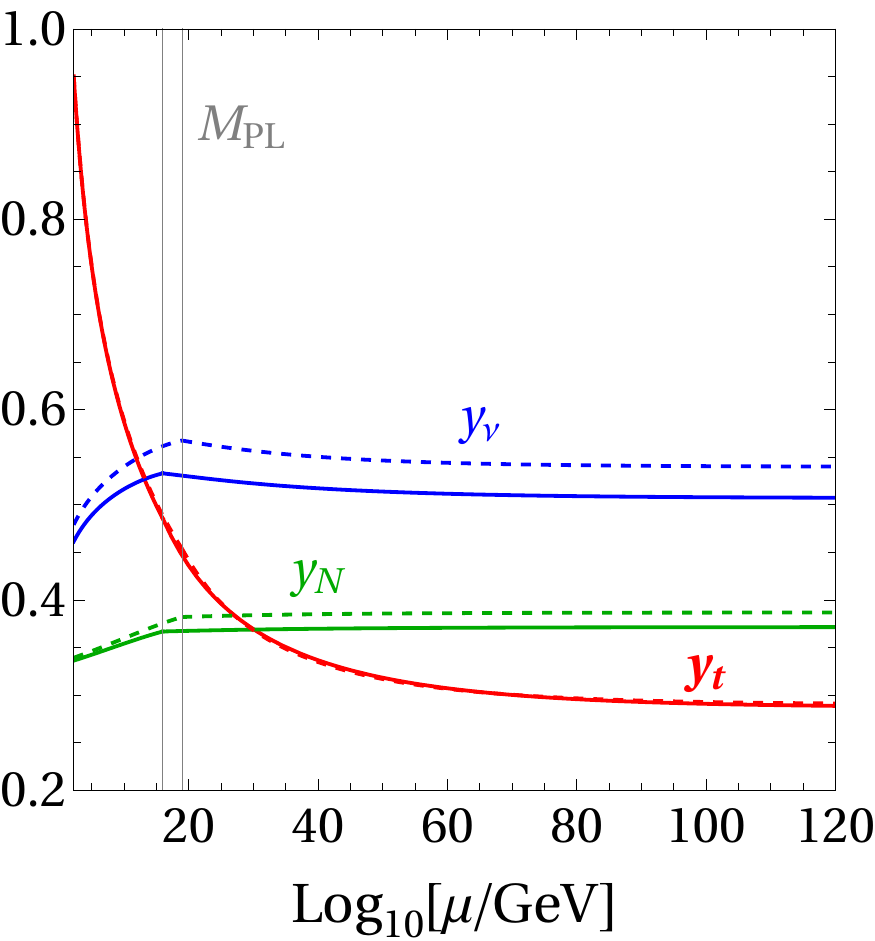}}\\
    \hspace{1cm}
    \subfloat[]{%
		\includegraphics[width=0.4\textwidth]{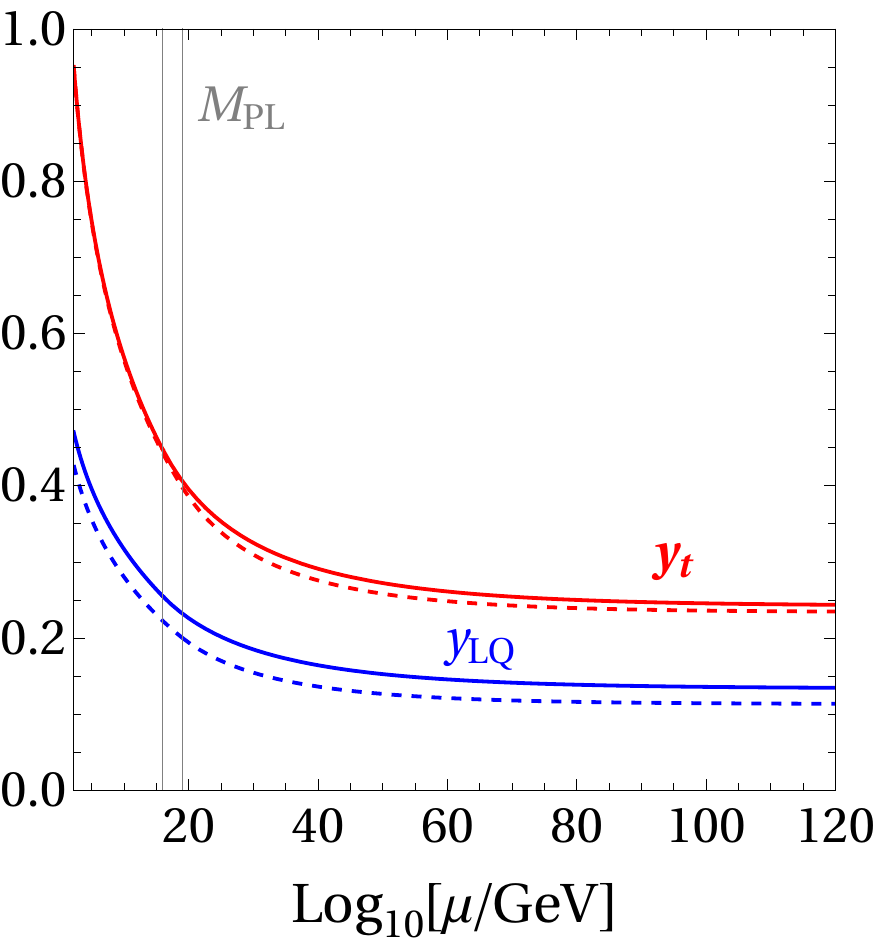}
		}   
\caption{(a) 1-loop RG flow of the hypercharge (red), dark gauge (blue), and kinetic mixing (green) couplings of the $B-L$ model with the Planck scale set at $10^{19}\gev$ (dashed) and $10^{16}\gev$ (solid). The gray vertical line indicates the fixed positions of the Planck scale. (b) Same as in (a) for the top Yukawa (red), the Dirac neutrino Yukawa (blue) and sterile neutrino Yukawa coupling (green) of the $B-L$ model.  (c) Same as in (a) for the top Yukawa (red) and LQ Yukawa coupling (blue) of the $S_3$ model.}
\label{fig:unc_mpl}
\end{figure}

The RG-invariance of the gauge-coupling ratios $r^\ast_{g,i}$ is made transparent in \reffig{fig:unc_mpl}(a), where we show the difference in the flow of the gauge couplings of the $B-L$ model for different choices of the Planck scale. The shift in the flow of the Yukawa sector of the $B-L$ models is presented in \reffig{fig:unc_mpl}(b), whereas the difference in the flow of the $S_3$ LQ model is shown in
\reffig{fig:unc_mpl}(c).

As was discussed in Sec.~\ref{sec:higher}, we observe focusing of the RG trajectories of the Yukawa couplings below the Planck scale. This indicates that the uncertainty calculated at the fixed point according to \refeq{eq:r2unc} is the maximal uncertainty that can be observed in a given NP model. 

\subsection{Scale-dependence of the gravitational corrections}

Finally, in this subsection we are going to drop the third of the approximations listed in Sec.~\ref{sec:intro}, \textit{i.e.}, we include in the analysis the scale dependence of the gravitational parameters. The actual functional form of $f_g$ and $f_y$ is subject to technical assumptions within the FRG framework. For example, in the Einstein-Hilbert truncation and in the Landau-gauge limit the relevant formulae can be found in Ref.\cite{Eichhorn:2017ylw}. In the following analysis we will avoid sticking to any particular form of $f_g(t)$ and $f_y(t)$ to be able to draw conclusions that remain as generic as possible. 

\paragraph{Gauge couplings}  Consider \refeq{eq:betag}, applied \textit{e.g.} to the three (irrelevant) gauge couplings of the $B-L$ model, $i=Y,d,\epsilon$. Due to its universality, the explicit gravitational contribution to the matter beta function $f_g$ entirely factors out of the running ratios:
\bea
\frac{d}{dt}\left(\frac{g_d}{g_Y}\right)&=&\frac{1}{g_Y}\left(\beta_d^{\textrm{matter}} -\frac{g_d}{g_Y}\beta_Y^{\textrm{matter}}\right)\left[t \right]\, \equiv \, F_d (g_Y(t),g_d(t),g_{\epsilon}(t),...)\,,\label{eq:rungratio1}\\
\frac{d}{dt}\left(\frac{g_\epsilon}{g_Y}\right)&=&\frac{1}{g_Y}\left(\beta_{\epsilon}^{\textrm{matter}} -\frac{g_\epsilon}{g_Y}\beta_Y^{\textrm{matter}} \right)\left[t \right]\,\equiv \,F_\epsilon(g_Y (t),g_d(t),g_{\epsilon}(t),...)\,,\label{eq:rungratio2}
\eea
where the matter beta functions $\beta_i^{\textrm{matter}}$ are given in parametric form in Eqs.~\eqref{eq:gyRGEKM}--\eqref{eq:geRGEKM} and we have formally defined slope functions $F_d$ and $F_{\epsilon}$ which do not depend on $f_g$. We are going to show that the 1-loop ratios $g_d/g_Y$ and $g_\epsilon/g_Y$ are exact invariants of the RG flow, whereas, at order $n\geq 2$, RG invariance is respected up to a very good approximation. 

To understand the underlying mechanism, let us apply the definition of total derivative to \refeq{eq:rungratio1} 
and focus on a sequence of infinitesimal scale intervals,
$...\,\, t_2< t_1 < t_0$, with
the system lying at the fixed point at $t_0$. Moving backwards in $t$, one gets
\bea
\frac{g_d(t_1)}{g_Y(t_1)}&=&r_{g,d}^{\ast}+(t_1-t_0)F_d(g_Y^\ast, g_d^{\ast}, g_{\epsilon}^{\ast},...)\label{eq:step1} \\
\frac{g_d(t_2)}{g_Y(t_2)}&=&\frac{g_d(t_1)}{g_Y(t_1)}+(t_2-t_1)F_d(g_Y(t_1), g_d(t_1), g_{\epsilon}(t_1),...) \\
\frac{g_d(t_3)}{g_Y(t_3)} & ... & \nonumber 
\eea
where $r_{g,d}^{\ast}$ was defined in \refeq{eq:gXFP} for REGs at arbitrary loop order. 
The same steps can be repeated for \refeq{eq:rungratio2}.
One can check, by directly imposing boundary condition~\eqref{eq:gXFP} into \refeq{eq:rungratio1} at $t_0$, 
that $F_d$ vanishes at the fixed point: $F_d(g_Y^\ast,...)=0$. Equivalently, 
one can show that $F_{\epsilon}(g_Y^\ast,...)=0$. Thus, $g_d(t_1)/g_Y(t_1)=r_{g,d}^{\ast}$ and $g_\epsilon(t_1)/g_Y(t_1)=r_{g,\epsilon}^{\ast}$. For the next time interval 
one can plug Eqs.~\eqref{eq:gXFP}, \eqref{eq:gEFP} into
Eqs.~\eqref{eq:gyRGEKM}--\eqref{eq:geRGEKM} and express 
the slope functions 
in terms of fixed-point ratios:
\be
F_d (g_Y(t_1),g_d(t_1),...)=\frac{g_Y^2(t_1)}{16\pi^2}\,\left[\tilde{b}_Y(t_1)r_{g,\epsilon}^{\ast\,2}+\tilde{b}_d(t_1)
r_{g,d}^{\ast\,2}+\tilde{b}_\epsilon(t_1) r_{g,\epsilon}^{\ast} r_{g,d}^{\ast}-\tilde{b}_Y(t_1)\right]r_{g,d}^{\ast}\,,\label{eq:fx1loop}
\ee
\be
F_{\epsilon} (g_Y(t_1),g_d(t_1),...)=\frac{g_Y^2(t_1)}{16\pi^2}\,\left[\left(\tilde{b}_Y(t_1) r_{g,\epsilon}^{\ast}+\tilde{b}_\epsilon(t_1) r_{g,d}^{\ast}\right)\left(1+r_{g,\epsilon}^{\ast\,2}\right)+\tilde{b}_d(t_1)\,r_{g,d}^{\ast\,2}\, r_{g,\epsilon}^{\ast}\right],\label{eq:feps1loop}
\ee
where, in agreement with \refeq{eq:loopcoe}, we have defined
\be\label{eq:loopcoet}
\tilde{b}_i(t) \equiv b_i+\Pi_{n\geq 2}^{(i)}(t)
\ee
away from the fixed point. 

At order $n=1$, $\tilde{b}_i(t)=b_i$. Boundary conditions~\eqref{eq:gXFP} and \eqref{eq:gEFP} ensure that
$F_d^{(n=1)}(t_1)=0$ and $F_{\epsilon}^{(n=1)} (t_1)=0$, 
independently of the actual values of the running gauge 
couplings. As an immediate consequence, $g_d(t)/g_Y(t)=r_{g,d}^{\ast}$ and $g_\epsilon(t)/g_Y(t)=r_{g,\epsilon}^{\ast}$ along 
the entire RG flow. Given that the ratios $r_{g,d}^{\ast}$ and $r_{g,\epsilon}^{\ast}$ are, at 1 loop, uniquely determined by the gauge quantum numbers, we can conclude that the predictions of AS for the gauge couplings do not depend on the particular functional 
form of the gravitational contribution $f_g (t)$.

On the other hand, the scale-invariance of the gauge coupling ratios 
is washed out by higher-order loop contributions, 
which introduce a $t$-dependent deviation from the 1-loop prediction due to appearance in \refeq{eq:loopcoet} 
of couplings (relevant gauge couplings, Yukawa and quartic couplings) that lie outside of the system \eqref{eq:gyRGEKM}--\eqref{eq:geRGEKM}. 
While these deviations from scale-invariance could potentially build up over scales spanning many orders of magnitude, they remain in practice quite limited.

\begin{figure}[t]
	\centering%
 		\includegraphics[width=0.4\textwidth]{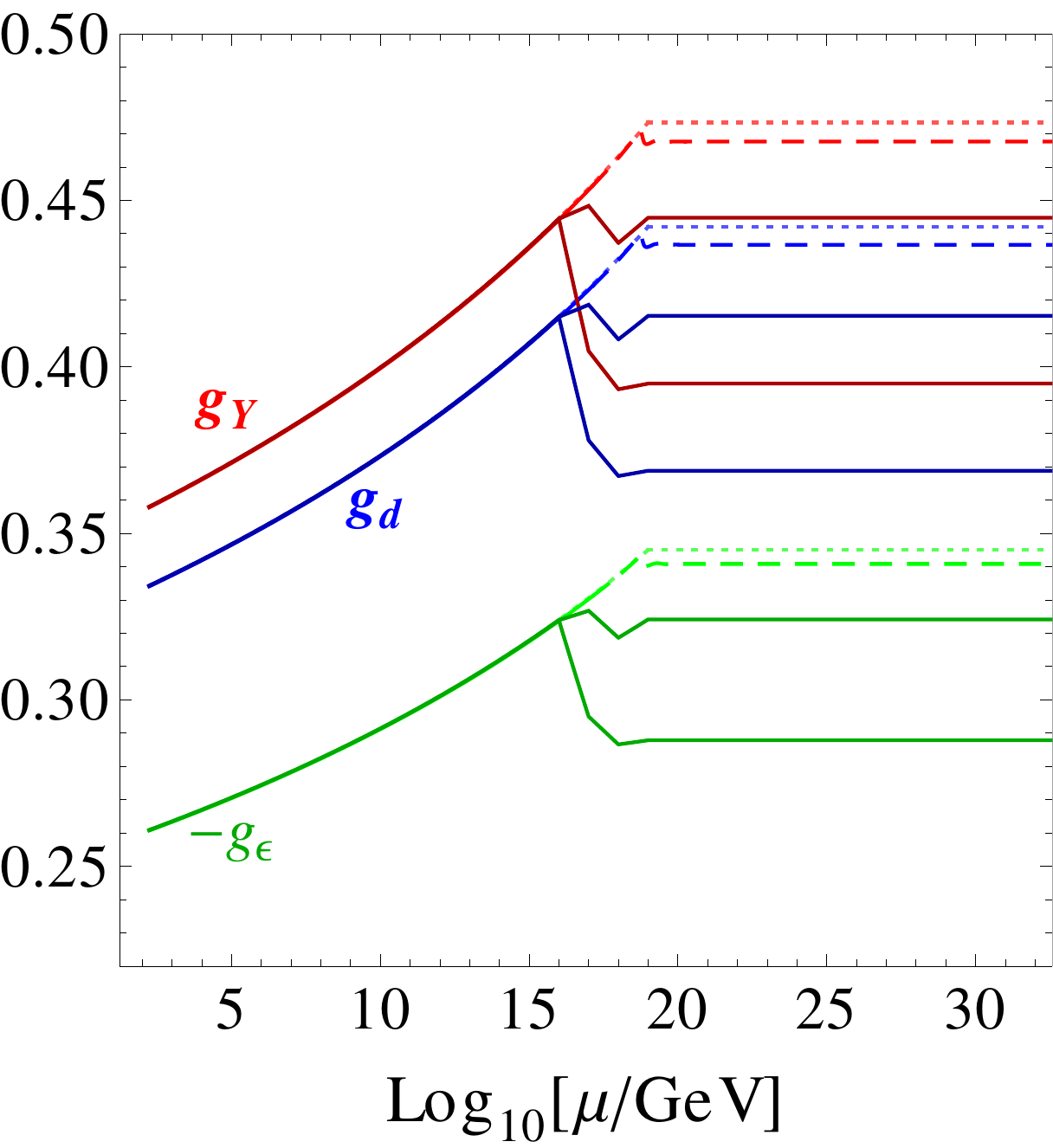}
\caption{RG flow of the hypercharge (red), dark gauge (blue), and kinetic mixing (green) couplings of the $B-L$ model at 1~loop. Dotted lines correspond to the benchmark scenario with $f_g$ constant above the Planck scale. Darker solid lines indicate two arbitrary parametrizations of $f_g(t)$ . Dashed lines correspond to the FRG results of Ref.\cite{Eichhorn:2017ylw}.}
\label{fig:unc_run}
\end{figure}

For illustration, we show in \reffig{fig:unc_run} the RG flow of the three gauge couplings of the $B-L$ model: $g_Y$ (red), $g_d$ (blue), and $-g_\epsilon$ (green). Dotted lines correspond to the benchmark scenario with $f_g$ and $f_y$ constant above the Planck scale ($\mpl=10^{19}\gev$). Darker solid lines indicate two random parametrizations of the functional dependence $f_g(t)$, where the gravity parameter is allow to vary by a factor~10 
in the range between $10^{16}\gev$ and $10^{20}\gev$. One can see that, in spite of different fixed-point values in each case, the RG invariance of the coupling ratios leads to unchanged low-scale predictions.
Finally, the dashed lines correspond to the $f_g(t)$ parametrization based on the FRG results of Ref.\cite{Eichhorn:2017ylw}. Note that in that framework it is reasonable to expect only very moderate changes in the gravity parameters.

\paragraph{Yukawa couplings} In the case of the Yukawa coupling RGEs
given in \refeq{eq:betay} 
one can consider the ratio of two irrelevant Yukawa couplings $y_1$ and $y_2$:
\bea\label{eq:runyratio}
\frac{d}{dt}\left(\frac{y_2}{y_1}\right)&=&\frac{1}{y_1}\left(\beta_2^{\textrm{matter}} -\frac{y_2}{y_1}\beta_1^{\textrm{matter}}\right)\left[t \right]\, \equiv \, G (y_j(t),g_k(t),...)\,,\label{eq:runyratio}
\eea
where the parametric form of the matter beta functions is given in \refeq{eq:yrRGE} and the slope function $G$ does not depend explicitly on $f_y$. Unlike the gauge sector, which features scale-invariant ratios at 1 loop, for the Yukawa couplings the $t$-dependence of $f_g$ and $f_y$ 
can impact the running of the $y_2/y_1$ ratio even at 1 loop. 

Repeating the infinitesimal-step analysis, 
\bea
\frac{y_2(t_1)}{y_1(t_1)}&=&\frac{y_2^\ast}{y_1^\ast}+(t_1-t_0)\,G(y_j^\ast, g_k^\ast,...)\,,\\
\frac{y_2(t_2)}{y_1(t_2)}&=&\frac{y_2(t_1)}{y_1(t_1)}+(t_2-t_1)\,G(y_j(t_1),g_k(t_1),...)\,,\\
\dots && \nonumber
\eea
one finds, again, that $G(y_j^\ast, g_k^\ast,...)=0$, so that $y_2(t_1)/y_1(t_1)=y_2^{\ast}/y_1^{\ast}$. However,
even at 1 loop, 
$G(y_j(t_1), g_k(t_1),...)\neq 0$. 

In order to see this, let us neglect for simplicity the additive contribution to the Yukawa beta functions in \refeq{eq:yrRGE}. In analogy to the gauge coupling case, one can then express
\begin{multline}\label{eq:Gyuk}
G(y_j(t_1), g_k(t_1),...)=\frac{y_1^2(t_1)\,r_{y,2}^{\ast}}{16\pi^2} \left[\sum_j \left(a_j^{(2)}-a_j^{(1)}\right) \frac{y_j^2(t_1)}{y_1^2(t_1)}- \sum_{l,k} \left(a_{lk}^{\prime (2)}-a_{lk}^{\prime (1)}\right) \frac{g_l(t_1)g_k(t_1)}{y_1^2(t_1)} \right]\\ 
+\frac{r_{y,2}^{\ast}}{16\pi^2} \sum_{n\geq 2}\left[\widetilde{\Pi}_n^{(2)}(t_1)-\widetilde{\Pi}_n^{(1)}(t_1) \right].  
\end{multline}
where $r_{y,2}^{\ast}$ was defined before \refeq{eq:delrgMPL}. Note that the non-abelian gauge couplings and some of the Yukawa couplings contributing to the sums within square brackets in \refeq{eq:Gyuk} correspond to relevant directions of the parameter space. Thus, even at 1 loop, if any of the couplings deviates from its fixed point (due to the change of the gravity parameters or the presence of relevant directions in the coupling space) the contribution from the first line of \refeq{eq:Gyuk} does not vanish exactly at $t_1$ and the ratio $y_2(t)/y_1(t)$ starts to flow. 

 \begin{figure}[t]
	\centering%
       \subfloat[]{%
 		\includegraphics[width=0.475\textwidth]{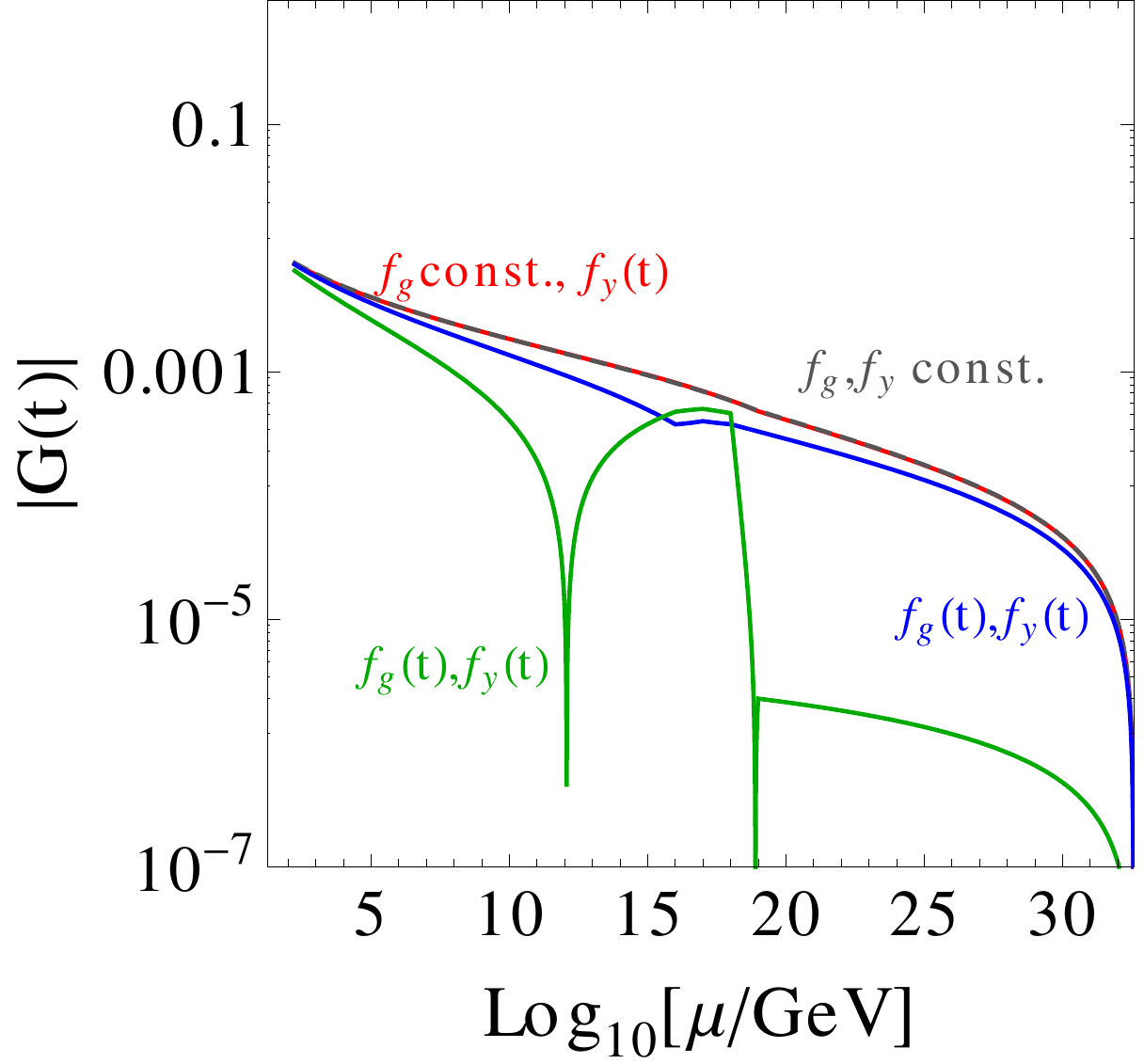}}
   \hspace{1cm}
          \subfloat[]{%
 		\includegraphics[width=0.4\textwidth]{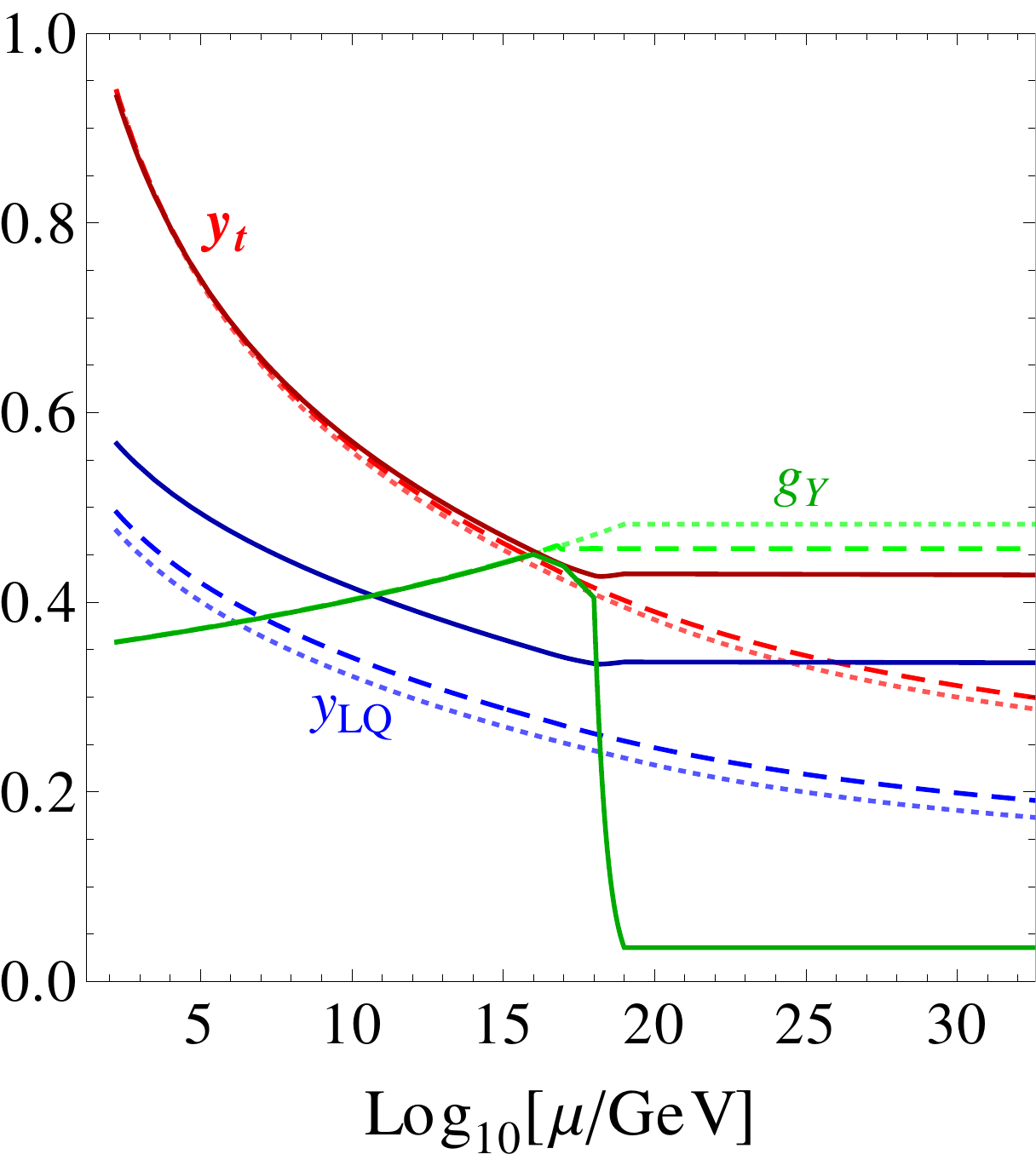}}
\caption{(a) Function $|G(t)|$ in the $S_3$ LQ model for different parametrizations of $f_g(t)$ and $f_y(t)$. Gray dashed line indicates the benchmark scenario with both gravitational parameters constant above the Planck scale ($\mpl=10^{19}\gev$). Red solid line corresponds to $f_g$ constant and $f_y$ allowed to vary by a factor $10$ in the range between $10^{16}\gev$ and $10^{20}\gev$. Blue and green lines show two arbitrary parametrizations of $f_g(t)$ and $f_y(t)$. (b) 1-loop RG flow of the top Yukawa (red), LQ Yukawa (blue) and hypercharge gauge (green) couplings of the $S_3$ model for different parametrizations of $f_g(t)$ and $f_y(t)$.  Dotted lines correspond to the benchmark 1-loop scenario with $f_g$ constant above the Planck scale. Dashed lines indicate a parametrization based on the FRG results of Ref.\cite{Eichhorn:2017ylw}. Solid lines show a parametrization resulting in $f_g(t)\ll |f_y(t)|$ in the trans-Planckian regime.}
\label{fig:runningG}
\end{figure}

This effect is illustrated in \reffig{fig:runningG}(a), where we show the actual size of the function $|G(t)|$ for different parametrizations of $f_g(t)$ and $f_y(t)$ in the $S_3$ LQ model. Gray dashed line indicates our benchmark scenario with both gravitational parameters constant above the Planck scale ($\mpl=10^{19}\gev$). Red solid line corresponds to a situation in which $f_g$ is kept constant, but $f_y$ is allowed to vary by a factor~10 in the range between $10^{16}\gev$ and $10^{20}\gev$. Once can see that there is no difference with the previous case, confirming the fact that the $f_y$ dependence factors out from the RG running of the Yukawa-coupling ratio. 
Finally, in green and blue we show two arbitrary parametrizations of $f_g(t)$, $f_y(t)$. Despite broad fluctuations, $|G(t)|$ remains small in size.
We can conclude that the flow of the ratio $y_2(t)/y_1(t)$ remains fairly stable throughout. 

As a matter of fact, the dominant source of uncertainty for the prediction of a Yukawa coupling 
is not given by the changing of  
$y_2(t)/y_1(t)$ along the RG flow, but rather by the fact that the fixed-point ratio $r_{y,2}^{\ast}$ 
itself becomes unknown once we factor in the $t$-dependence of $f_g$ and $f_y$. As was shown in several papers\cite{Alkofer:2020vtb,Kowalska:2020gie,Kowalska:2020zve}, at 1 loop in the matter RGEs the fixed-point ratio 
of two Yukawa couplings takes the parametric form
\be\label{eq:ratiofgfy}
r_{y,2}^\ast=\sqrt{\frac{A'\,f_g+B'\,f_y}{A\,f_g+B\,f_y}}\,.
\ee
Numerical coefficients $A$, $A'$, $B$, $B'$ 
are combinations of the 1-loop coefficients of the matter beta functions. They can be finite or zero. 
Referring, again, to the 
simple case of two irrelevant Yukawa couplings, $y_1$, $y_2$, 
and one gauge coupling, $g_1$, one can write
\bea
A&=&a_2^{(2)}a_{11}^{\prime(1)}-a_2^{(1)}a_{11}^{\prime(2)},\qquad B=b_1\left(a_2^{(2)}-a_2^{(1)}\right)\,,\\
A'&=&a_1^{(1)}a_{11}^{\prime(2)}-a_1^{(2)}a_{11}^{\prime(1)},\qquad B'=b_1\left(a_1^{(1)}-a_1^{(2)}\right)\,,
\eea
where $a_{j}^{(r)}$, $a_{lk}^{\prime (r)}$ are coefficients of the Yukawa-coupling beta functions, cf.~\refeq{eq:yrRGE},
and $b_1$ is the 1-loop coefficient of the $g_1$ beta function. 

In the limiting case of one of the two gravitational parameters being strongly dominant with respect 
to the other, the ratio becomes known, albeit possibly extreme (even zero or infinity). For example, in 
the $S_3$ LQ model one gets
\be
\lim_{f_g\to \infty} r_{y,2}^\ast=\sqrt{\frac{A'}{A}}=0.55,\qquad\textrm{or}\qquad
\lim_{|f_y|\to \infty}r_{y,2}^\ast=\sqrt{\frac{B'}{B}}=0.78\,,
\ee
which are not far from the prediction obtained from \refeq{eq:lep1_fp}, $r_{y,2}^{\ast}\approx 0.48$.
In general, however, cancellations in the numerator/denominator of \refeq{eq:ratiofgfy} might take place, so that there is no real handle on the prediction.   

This is illustrated in \reffig{fig:runningG}(b), where we show the RG flow of the top Yukawa (red), LQ Yukawa (blue) and hypercharge (green) gauge coupling of the $S_3$ model with different functional forms of $f_g(t)$ and $f_y(t)$. The color code is the same as in \reffig{fig:unc_run}(a). While it was shown above that even extreme fluctuations do not modify the low-scale predictions for the gauge couplings, we can see in \reffig{fig:runningG}(b) that the trans-Planckian behavior of the running
$y_{\textrm{LQ}}$~(blue) and $y_t$~(red) couplings is drastically different in the cases 
in solid vs.~those in dotted/dashed. 

An assessment from first principles of the 
uncertainties on the action 
was attempted in Ref.\cite{Dona:2013qba}. Several choices of the truncation, and different matter content, were shown to induce significant shifts in the fixed-point values of the cosmological and Newton constant. On the other hand, when those shifts are plugged into typical FRG computations of $f_g$ and $f_y$ -- \textit{e.g.}, those in Ref.\cite{Eichhorn:2017ylw} -- the actual percent uncertainty on $f_g(t)$ and $f_y(t)$ rarely exceeds the $\mathcal{O}(1)$ level along the entire trans-Planckian flow. In \reffig{fig:runningG}(b), the parametrization based on the explicit 
FRG results of Ref.\cite{Eichhorn:2017ylw} is depicted in dashed lines. It introduces a small 
uncertainty on the prediction of $y_{\textrm{LQ}}$, of around $5\%$. 

Incidentally, in the sub-Planckian regime of \reffig{fig:runningG}(b) we observe focusing due to the presence of a large coupling $g_3$, in agreement with the discussion at the very end of Sec.~\ref{sec:higher}. The resulting low-scale uncertainty on $y_{\textrm{LQ}}$ reduces to $4\%$ for the FRG-inspired functional dependence of the gravitational parameters. For the case in solid, which corresponds to $f_g(t)\ll |f_y(t)|$,
the low-scale uncertainty on $y_{\textrm{LQ}}$ reads $\sim19\%$.


\subsection{Relevant abelian gauge couplings\label{sec:abel_gau}}

The systematic analysis featured in the three subsections presented above relies on the implicit choice of a maximally predictive fixed point for the particle physics plus gravity system. This was guaranteed by enforcing assumptions~1, 2, and 3 in Sec.~\ref{sec:gen_not}. 

As was mentioned at the beginning of Sec.~\ref{sec:higher}, however, often one may be interested in obtaining predictions for irrelevant Yukawa couplings in a scenario where all the gauge couplings, including the abelian ones, correspond to relevant directions around a Gaussian fixed point -- $g_Y^{\ast}=0$, etc. All gauge couplings become, in other words, asymptotically free like the non-abelian ones. Such situation emerges quite naturally if $f_g$ is large enough (larger, for example, than the values given in \refeq{eq:fgBML} and \refeq{eq:fgS3} for the $B-L$ and $S_3$ LQ models, respectively) and it corresponds, effectively, to relaxing assumption~1 in Sec.~\ref{sec:gen_not}. 

We discuss in this subsection what uncertainties we expect in this case for the predicted Yukawa couplings.  
When $g_Y^\ast=0$, the irrelevant fixed points of the Yukawa-coupling system do not depend on the value of $f_g$. Consequently, the dependence on $f_y$ factors out of the ratio $r_{y,2}^{\ast}$, which then becomes fully determined by the  1-loop coefficients of the beta functions, see \refeq{eq:ratiofgfy}. That does not mean, however, that predictivity in the Yukawa sector is fully restored. In fact, the presence of relevant parameters typically affects the RGEs of different Yukawa couplings unequally. 

Let us consider two arbitrary values $f_g,f'_g>f_g^{\textrm{AS}}$, where $f_g^{\textrm{AS}}$ would be the value 
making the fixed point in $g_Y^{\ast}$ irrelevant -- for example, the value given in \refeq{eq:fgBML} or the one 
in \refeq{eq:fgS3}. Unlike $f_g^{\textrm{AS}}$, $f_g$ and $f'_g$ cannot be predicted 
from low-energy measurements and are thus unknown. Given  
the beta-function system of \refeq{eq:yrRGE} (at 1-loop order) and 
following the steps that led to \refeq{eq:yukrat}, we can derive the shift on the Yukawa-coupling fixed-point prediction computed with $f'_g$ with respect to $f_g$:
\be
y_2^{\ast}(f'_g)= \left[F_1\left(a_j^{(r)}\right)\left( y_1^{\ast 2}(f_g)+\delta y_1^{\ast 2} \right)\right]^{1/2},
\ee
where 
\be
\delta y_1^{\ast 2}= y_1^{\ast 2}(f'_g)-y_1^{\ast 2}(f_g)\,.
\ee
Note that, unlike in \refeq{eq:2lshift}, the fixed-point values cannot be here controlled without explicit knowledge of the quantum gravity parameters. 

To get a quantitative sense of the shift observed in predictions, 
let us focus on the  
$S_3$ LQ model and consider two arbitrary choices leading to relevant fixed points in the gauge-coupling system: $f_g=0.0107$ and $f'_g=0.0212$. Recalling that $y_2^{\ast}\equiv y_{\textrm{LQ}}^{\ast}$, the relative uncertainty reads
\be
\frac{y_{\textrm{LQ}}^{\ast}(f'_g)-y_{\textrm{LQ}}^{\ast}(f_g)}{y_{\textrm{LQ}}^{\ast}(f_g)}\approx 68\%\,,
\ee
approximately of the size of the relative uncertainty on the parameter $f_g$ itself. We observe similar numerical values for the predictions of the $B-L$ model. 

On the other hand, one should keep in mind that, notwithstanding
what the actual numerical value of $f_g > f_g^{\textrm{AS}}$ may be, it should always lead the running gauge couplings $g_3$, $g_2$, and $g_Y$ to match their low-scale determination (which is independent of $f_g$). As such, we expect the predictions in the Yukawa sector to be subject to significant focusing in their flow 
from the fixed point down to the EWSB scale. In fact, in the case of the $S_3$ LQ we observe a significant reduction of the uncertainty:
\be
\frac{y_{\textrm{LQ}}(M_t,f'_g)-y_{\textrm{LQ}}(M_t,f_g)}{y_{\textrm{LQ}}(M_t,f_g)}\approx 6\%\,.
\ee
As was the case in Sec.~\ref{sec:higher}, we can state a general rule of thumb, that the uncertainty at the fixed point is always larger that the actual one at the low-energy scale. 
\section{Conclusions\label{sec:summary}}

In this paper, we have investigated in detail the issue of the theoretical uncertainties associated with predictions for
irrelevant Lagrangian couplings 
emerging from trans-Planckian AS. We have chosen 
a complementary approach with respect to Ref.\cite{Dona:2013qba}: while that article evaluated the uncertainties of the quantum-gravity sector in the context of calculations within the FRG, in this work we have decided to bypass strictly gravitational aspects following a popular phenomenological approach which relies on a parametric 
description of quantum gravity with universal coefficients that are eventually obtained from low-energy observations. Our focus lies squarely on the uncertainties pertaining to the matter sector, which takes the form of a system of gauge- and Yukawa-coupling RGEs. 

We have quantified the impact of relaxing several approximations commonly used in the literature to extract phenomenological predictions. In particular we have considered: the effect of including higher-order corrections in the matter RGEs of the gauge-Yukawa system; the impact of selecting a value different from $10^{19}\gev$ for the (somewhat arbitrary) position of the Planck scale; and the effects 
of using scale-dependent parametrizations of the gravitational UV completion in the matter RGEs, resulting in a non-instantaneous decoupling of gravity from matter around the Planck scale. 

The main findings of our study are the following. First, in the gauge sector the uncertainty induced by relaxing any of the simplifying assumptions never exceeds the $1\%$ level. We can conclude that fixed-point predictions for the irrelevant gauge couplings of the SM and/or NP models are extremely robust, even when they are obtained in a heuristic, simplified approach to AS that is based on some approximations. 

A similar conclusion can be drawn for the Yukawa sector of the NP theory, if the predicted Yukawa couplings are of comparable size to the irrelevant gauge couplings. The uncertainties remain at bay, not exceeding $\sim 10\%$ at the fixed point if higher-order corrections are included, or if the Planck scale is moved to, \textit{e.g.}, $10^{16}\gev$. The situation is additionally helped by focusing of the RG trajectories in the sub-Planckian regime.  

Potentially more dangerous uncertainties could stem from considering the non-trivial scale dependence of the gravitational contributions to the matter beta functions, parameterized by functions $f_g(t)$ and $f_y(t)$, as in this case we lose the ability of determining the actual value of the Yukawa couplings at the fixed point. However, we have argued, based on both an analytical and numerical discussion, that in the range of variability of the gravitation parameters that can be realistically expected in the framework of the FRG, the resulting uncertainty is moderate. 

Finally, we have identified one situation in which the AS-based predictions cannot be trusted as any modification of the setup would lead to a drastic change on the predicted value of a NP coupling. This happens if the predicted Yukawa coupling is much smaller than the irrelevant gauge couplings, since in this case it is a result of an accidental, precise cancellation of two quantities of comparable size. This is a manifestation of fine tuning, exactly analogous to 
some of the cases described in, \textit{e.g.}, 
Ref.\cite{Kowalska:2022ypk}, in which the gravitational parameter $f_y$ happened to lie unnaturally close to a critical value $f_{Z,XY}^{\textrm{crit}}$ which depends exclusively on the gauge quantum numbers of the matter sector. Since the prediction then finds its origin on a fortuitous cancellation of unrelated quantities it is very subject to the approximations employed for its derivation.  

\section*{Acknowledgements}    

KK and EMS would like to thank the Leinweber Center for Theoretical Physics at the University of Michigan for the kind hospitality during the initial stages of this work.
WK and EMS are supported by the National Science Centre (Poland) under the research grant 2020/\allowbreak38/\allowbreak E/\allowbreak ST2/\allowbreak00126.
KK and DR are supported by the National Science Centre (Poland) under the research grant 2017/\allowbreak26/\allowbreak E/\allowbreak ST2/\allowbreak 00470. The use of the CIS computer cluster at the National Centre for Nuclear Research in Warsaw is gratefully
acknowledged.

\appendix
\section{Renormalization group equations}
\label{app:rges}

\newcommand{\tr}{\mathrm{Tr}}
\newcommand{\trans}{\mathrm{T}}

In this appendix we presents the DREG matter RGEs used in this work. They are listed order-by-order according to the following notation:
\begin{equation*}    
\beta\left(X\right) \equiv \mu \frac{d X}{d \mu}\equiv \frac{1}{\left(4 \pi\right)^{2}}\beta^{(1)}(X)+ \frac{1}{\left(4 \pi\right)^{4}}\beta^{(2)}(X)\,.
\end{equation*}
We do not write explicitly the contributions from the down-type quark and charged-lepton Yukawa matrices which do not affect our numerical analysis. 

\subsection{Gauged U(1)$_{B-L}$}
\subsubsection{Gauge sector}

\begin{align*}
\begin{autobreak}
\beta^{(1)}(g_{Y}) =\frac{41}{6} g_{Y}^{3}
\end{autobreak}
\end{align*}
\begin{align*}
\begin{autobreak}
\beta^{(2)}(g_{Y}) =

+ \frac{199}{18} g_{Y}^{5}

+ \frac{92}{9} g_{Y}^{3} g_{d}^{2}

+ \frac{199}{18} g_{Y}^{3} g_\epsilon^{2}

+ \frac{164}{9} g_{Y}^{3} g_{d} g_\epsilon

+ \frac{9}{2} g_2^{2} g_{Y}^{3}

+ \frac{44}{3} g_3^{2} g_{Y}^{3}

-  \frac{17}{6} g_{Y}^{3} \tr\left(Y_u^{\dagger} Y_u \right)

-  \frac{1}{2} g_{Y}^{3} \tr\left(Y_\nu^{\dagger} Y_\nu \right)
\end{autobreak}
\end{align*}

\begin{align*}
\begin{autobreak}
\beta^{(1)}(g_\epsilon) =

+ \frac{32}{3} g_{Y}^{2} g_{d}

+ \frac{32}{3} g_{d} g_\epsilon^{2}

+ \frac{41}{3} g_{Y}^{2} g_\epsilon

+ 12 g_{d}^{2} g_\epsilon

+ \frac{41}{6} g_\epsilon^{3}
\end{autobreak}
\end{align*}
\begin{align*}
\begin{autobreak}
\beta^{(2)}(g_\epsilon) =

+ \frac{224}{9} g_{Y}^{2} g_{d}^{3}

+ \frac{199}{6} g_{Y}^{2} g_\epsilon^{3}

+ \frac{164}{9} g_{Y}^{4} g_{d}

+ \frac{656}{9} g_{Y}^{2} g_{d} g_\epsilon^{2}

+ \frac{448}{9} g_{d}^{3} g_\epsilon^{2}

+ \frac{184}{3} g_{d}^{2} g_\epsilon^{3}

+ \frac{328}{9} g_{d} g_\epsilon^{4}

+ 12 g_2^{2} g_{Y}^{2} g_{d}

+ 12 g_2^{2} g_{d} g_\epsilon^{2}

+ \frac{32}{3} g_3^{2} g_{Y}^{2} g_{d}

+ \frac{32}{3} g_3^{2} g_{d} g_\epsilon^{2}

+ \frac{199}{9} g_{Y}^{4} g_\epsilon

+ \frac{644}{9} g_{Y}^{2} g_{d}^{2} g_\epsilon

+ \frac{800}{9} g_{d}^{4} g_\epsilon

+ \frac{199}{18} g_\epsilon^{5}

+ 9 g_2^{2} g_{Y}^{2} g_\epsilon

+ 12 g_2^{2} g_{d}^{2} g_\epsilon

+ \frac{88}{3} g_3^{2} g_{Y}^{2} g_\epsilon

+ \frac{32}{3} g_3^{2} g_{d}^{2} g_\epsilon

+ \frac{9}{2} g_2^{2} g_\epsilon^{3}

+ \frac{44}{3} g_3^{2} g_\epsilon^{3}

-  \frac{10}{3} g_{Y}^{2} g_{d} \tr\left(Y_u^{\dagger} Y_u \right)

- 2 g_{Y}^{2} g_{d} \tr\left(Y_\nu^{\dagger} Y_\nu \right)

-  \frac{10}{3} g_{d} g_\epsilon^{2} \tr\left(Y_u^{\dagger} Y_u \right)

- 2 g_{d} g_\epsilon^{2} \tr\left(Y_\nu^{\dagger} Y_\nu \right)

-  \frac{17}{3} g_{Y}^{2} g_\epsilon \tr\left(Y_u^{\dagger} Y_u \right)

-  g_{Y}^{2} g_\epsilon \tr\left(Y_\nu^{\dagger} Y_\nu \right)

-  \frac{4}{3} g_{d}^{2} g_\epsilon \tr\left(Y_u^{\dagger} Y_u \right)

- 4 g_{d}^{2} g_\epsilon \tr\left(Y_\nu^{\dagger} Y_\nu \right)

- 4 g_{d}^{2} g_\epsilon \tr\left(Y_N^{*} Y_N \right)

-  \frac{17}{6} g_\epsilon^{3} \tr\left(Y_u^{\dagger} Y_u \right)

-  \frac{1}{2} g_\epsilon^{3} \tr\left(Y_\nu^{\dagger} Y_\nu \right)
\end{autobreak}
\end{align*}

\begin{align*}
\begin{autobreak}
\beta^{(1)}(g_{d}) =

+ 12 g_{d}^{3}

+ \frac{41}{6} g_{d} g_\epsilon^{2}

+ \frac{32}{3} g_{d}^{2} g_\epsilon
\end{autobreak}
\end{align*}
\begin{align*}
\begin{autobreak}
\beta^{(2)}(g_{d}) =

+ \frac{92}{9} g_{Y}^{2} g_{d}^{3}

+ \frac{199}{18} g_{Y}^{2} g_{d} g_\epsilon^{2}

+ \frac{800}{9} g_{d}^{5}

+ \frac{184}{3} g_{d}^{3} g_\epsilon^{2}

+ \frac{328}{9} g_{d}^{2} g_\epsilon^{3}

+ \frac{199}{18} g_{d} g_\epsilon^{4}

+ \frac{9}{2} g_2^{2} g_{d} g_\epsilon^{2}

+ \frac{44}{3} g_3^{2} g_{d} g_\epsilon^{2}

+ \frac{164}{9} g_{Y}^{2} g_{d}^{2} g_\epsilon

+ \frac{448}{9} g_{d}^{4} g_\epsilon

+ 12 g_2^{2} g_{d}^{2} g_\epsilon

+ \frac{32}{3} g_3^{2} g_{d}^{2} g_\epsilon

+ 12 g_2^{2} g_{d}^{3}

+ \frac{32}{3} g_3^{2} g_{d}^{3}

-  \frac{4}{3} g_{d}^{3} \tr\left(Y_u^{\dagger} Y_u \right)

- 4 g_{d}^{3} \tr\left(Y_\nu^{\dagger} Y_\nu \right)

- 4 g_{d}^{3} \tr\left(Y_N^{*} Y_N \right)

-  \frac{17}{6} g_{d} g_\epsilon^{2} \tr\left(Y_u^{\dagger} Y_u \right)

-  \frac{1}{2} g_{d} g_\epsilon^{2} \tr\left(Y_\nu^{\dagger} Y_\nu \right)

-  \frac{10}{3} g_{d}^{2} g_\epsilon \tr\left(Y_u^{\dagger} Y_u \right)

- 2 g_{d}^{2} g_\epsilon \tr\left(Y_\nu^{\dagger} Y_\nu \right)
\end{autobreak}
\end{align*}

\begin{align*}
\begin{autobreak}
\beta^{(1)}(g_2) =- \frac{19}{6} g_2^{3}
\end{autobreak}
\end{align*}
\begin{align*}
\begin{autobreak}
\beta^{(2)}(g_2) =

+ 4 g_2^{3} g_{d} g_\epsilon

+ \frac{3}{2} g_2^{3} g_{Y}^{2}

+ 4 g_2^{3} g_{d}^{2}

+ \frac{3}{2} g_2^{3} g_\epsilon^{2}

+ \frac{35}{6} g_2^{5}

+ 12 g_2^{3} g_3^{2}

-  \frac{3}{2} g_2^{3} \tr\left(Y_u^{\dagger} Y_u \right)

-  \frac{1}{2} g_2^{3} \tr\left(Y_\nu^{\dagger} Y_\nu \right)
\end{autobreak}
\end{align*}

\begin{align*}
\begin{autobreak}
\beta^{(1)}(g_3) =-7 g_3^{3}
\end{autobreak}
\end{align*}
\begin{align*}
\begin{autobreak}
\beta^{(2)}(g_3) =

+ \frac{4}{3} g_3^{3} g_{d} g_\epsilon

+ \frac{9}{2} g_2^{2} g_3^{3}

+ \frac{11}{6} g_3^{3} g_{Y}^{2}

+ \frac{4}{3} g_3^{3} g_{d}^{2}

+ \frac{11}{6} g_3^{3} g_\epsilon^{2}

- 26 g_3^{5}

- 2 g_3^{3} \tr\left(Y_u^{\dagger} Y_u \right)
\end{autobreak}
\end{align*}

\subsubsection{Yukawa sector}
The Higgs potential couplings $\lambda_1$, $\lambda_2$, and $\lambda_3$ are defined according to Ref.\,\cite{Coriano:2015sea}.
\begin{align*}
\begin{autobreak}
\beta^{(1)}(Y_u) =

+ \frac{3}{2} Y_u Y_u^{\dagger} Y_u

+ 3 \tr\left(Y_u^{\dagger} Y_u \right) Y_u

+ \tr\left(Y_\nu^{\dagger} Y_\nu \right) Y_u

-  \frac{17}{12} g_{Y}^{2} Y_u

-  \frac{2}{3} g_{d}^{2} Y_u

-  \frac{5}{3} g_{d} g_\epsilon Y_u

-  \frac{17}{12} g_\epsilon^{2} Y_u

-  \frac{9}{4} g_2^{2} Y_u

- 8 g_3^{2} Y_u
\end{autobreak}
\end{align*}
\begin{align*}
\begin{autobreak}
\beta^{(2)}(Y_u) =

+ \frac{3}{2} Y_u Y_u^{\dagger} Y_u Y_u^{\dagger} Y_u

-  \frac{27}{4} \tr\left(Y_u^{\dagger} Y_u Y_u^{\dagger} Y_u \right) Y_u

-  \frac{27}{4} \tr\left(Y_u^{\dagger} Y_u \right) Y_u Y_u^{\dagger} Y_u

-  \frac{9}{4} \tr\left(Y_\nu^{\dagger} Y_\nu \right) Y_u Y_u^{\dagger} Y_u

-  \frac{9}{4} \tr\left(Y_\nu^{\dagger} Y_\nu Y_\nu^{\dagger} Y_\nu \right) Y_u

- 3 \tr\left(Y_\nu^{\dagger} Y_\nu Y_N^{*} Y_N \right) Y_u

- 12 \lambda_1 Y_u Y_u^{\dagger} Y_u

+ 6 \lambda_1^{2} Y_u

+ \frac{1}{2} \lambda_3^{2} Y_u

+ \frac{223}{48} g_{Y}^{2} Y_u Y_u^{\dagger} Y_u

+ \frac{4}{3} g_{d}^{2} Y_u Y_u^{\dagger} Y_u

+ \frac{25}{12} g_{d} g_\epsilon Y_u Y_u^{\dagger} Y_u

+ \frac{223}{48} g_\epsilon^{2} Y_u Y_u^{\dagger} Y_u

+ \frac{135}{16} g_2^{2} Y_u Y_u^{\dagger} Y_u

+ 16 g_3^{2} Y_u Y_u^{\dagger} Y_u

+ \frac{85}{24} g_{Y}^{2} \tr\left(Y_u^{\dagger} Y_u \right) Y_u

+ \frac{5}{3} g_{d}^{2} \tr\left(Y_u^{\dagger} Y_u \right) Y_u

+ \frac{25}{6} g_{d} g_\epsilon \tr\left(Y_u^{\dagger} Y_u \right) Y_u

+ \frac{85}{24} g_\epsilon^{2} \tr\left(Y_u^{\dagger} Y_u \right) Y_u

+ \frac{45}{8} g_2^{2} \tr\left(Y_u^{\dagger} Y_u \right) Y_u

+ 20 g_3^{2} \tr\left(Y_u^{\dagger} Y_u \right) Y_u

+ \frac{5}{8} g_{Y}^{2} \tr\left(Y_\nu^{\dagger} Y_\nu \right) Y_u

+ 5 g_{d}^{2} \tr\left(Y_\nu^{\dagger} Y_\nu \right) Y_u

+ \frac{5}{2} g_{d} g_\epsilon \tr\left(Y_\nu^{\dagger} Y_\nu \right) Y_u

+ \frac{5}{8} g_\epsilon^{2} \tr\left(Y_\nu^{\dagger} Y_\nu \right) Y_u

+ \frac{15}{8} g_2^{2} \tr\left(Y_\nu^{\dagger} Y_\nu \right) Y_u

+ \frac{1187}{216} g_{Y}^{4} Y_u

+ \frac{91}{12} g_{Y}^{2} g_{d}^{2} Y_u

+ \frac{1187}{108} g_{Y}^{2} g_\epsilon^{2} Y_u

+ \frac{203}{27} g_{d}^{4} Y_u

+ \frac{1085}{36} g_{d}^{2} g_\epsilon^{2} Y_u

+ \frac{502}{27} g_{Y}^{2} g_{d} g_\epsilon Y_u

+ \frac{502}{27} g_{d} g_\epsilon^{3} Y_u

+ \frac{9}{4} g_2^{2} g_{d} g_\epsilon Y_u

-  \frac{20}{9} g_3^{2} g_{d} g_\epsilon Y_u

+ \frac{665}{27} g_{d}^{3} g_\epsilon Y_u

+ \frac{1187}{216} g_\epsilon^{4} Y_u

-  \frac{3}{4} g_2^{2} g_{Y}^{2} Y_u

+ \frac{3}{4} g_2^{2} g_{d}^{2} Y_u

-  \frac{3}{4} g_2^{2} g_\epsilon^{2} Y_u

-  \frac{23}{4} g_2^{4} Y_u

+ 9 g_2^{2} g_3^{2} Y_u

+ \frac{19}{9} g_3^{2} g_{Y}^{2} Y_u

-  \frac{8}{9} g_3^{2} g_{d}^{2} Y_u

+ \frac{19}{9} g_3^{2} g_\epsilon^{2} Y_u

- 108 g_3^{4} Y_u
\end{autobreak}
\end{align*}

\begin{align*}
\begin{autobreak}
\beta^{(1)}(Y_\nu) =

+ \frac{3}{2} Y_\nu Y_\nu^{\dagger} Y_\nu

+ 2 Y_\nu Y_N^{*} Y_N

+ 3 \tr\left(Y_u^{\dagger} Y_u \right) Y_\nu

+ \tr\left(Y_\nu^{\dagger} Y_\nu \right) Y_\nu

-  \frac{3}{4} g_{Y}^{2} Y_\nu

- 6 g_{d}^{2} Y_\nu

- 3 g_{d} g_\epsilon Y_\nu

-  \frac{3}{4} g_\epsilon^{2} Y_\nu

-  \frac{9}{4} g_2^{2} Y_\nu
\end{autobreak}
\end{align*}
\begin{align*}
\begin{autobreak}
\beta^{(2)}(Y_\nu) =

+ \frac{3}{2} Y_\nu Y_\nu^{\dagger} Y_\nu Y_\nu^{\dagger} Y_\nu

+ 7 Y_\nu Y_N^{*} Y_\nu^{\trans} Y_\nu^{*} Y_N

-  \frac{1}{2} Y_\nu Y_N^{*} Y_N Y_\nu^{\dagger} Y_\nu

- 2 Y_\nu Y_N^{*} Y_N Y_N^{*} Y_N

-  \frac{27}{4} \tr\left(Y_u^{\dagger} Y_u Y_u^{\dagger} Y_u \right) Y_\nu

-  \frac{27}{4} \tr\left(Y_u^{\dagger} Y_u \right) Y_\nu Y_\nu^{\dagger} Y_\nu

-  \frac{9}{4} \tr\left(Y_\nu^{\dagger} Y_\nu Y_\nu^{\dagger} Y_\nu \right) Y_\nu

-  \frac{9}{4} \tr\left(Y_\nu^{\dagger} Y_\nu \right) Y_\nu Y_\nu^{\dagger} Y_\nu

- 3 \tr\left(Y_\nu^{\dagger} Y_\nu Y_N^{*} Y_N \right) Y_\nu

- 6 \tr\left(Y_N^{*} Y_N \right) Y_\nu Y_N^{*} Y_N

- 12 \lambda_1 Y_\nu Y_\nu^{\dagger} Y_\nu

- 8 \lambda_3 Y_\nu Y_N^{*} Y_N

+ 6 \lambda_1^{2} Y_\nu

+ \frac{1}{2} \lambda_3^{2} Y_\nu

+ \frac{93}{16} g_{Y}^{2} Y_\nu Y_\nu^{\dagger} Y_\nu

+ 12 g_{d}^{2} Y_\nu Y_\nu^{\dagger} Y_\nu

+ \frac{39}{4} g_{d} g_\epsilon Y_\nu Y_\nu^{\dagger} Y_\nu

+ \frac{93}{16} g_\epsilon^{2} Y_\nu Y_\nu^{\dagger} Y_\nu

+ \frac{135}{16} g_2^{2} Y_\nu Y_\nu^{\dagger} Y_\nu

+ 40 g_{d}^{2} Y_\nu Y_N^{*} Y_N

- 12 g_{d} g_\epsilon Y_\nu Y_N^{*} Y_N

+ \frac{85}{24} g_{Y}^{2} \tr\left(Y_u^{\dagger} Y_u \right) Y_\nu

+ \frac{5}{3} g_{d}^{2} \tr\left(Y_u^{\dagger} Y_u \right) Y_\nu

+ \frac{25}{6} g_{d} g_\epsilon \tr\left(Y_u^{\dagger} Y_u \right) Y_\nu

+ \frac{85}{24} g_\epsilon^{2} \tr\left(Y_u^{\dagger} Y_u \right) Y_\nu

+ \frac{45}{8} g_2^{2} \tr\left(Y_u^{\dagger} Y_u \right) Y_\nu

+ 20 g_3^{2} \tr\left(Y_u^{\dagger} Y_u \right) Y_\nu

+ \frac{5}{8} g_{Y}^{2} \tr\left(Y_\nu^{\dagger} Y_\nu \right) Y_\nu

+ 5 g_{d}^{2} \tr\left(Y_\nu^{\dagger} Y_\nu \right) Y_\nu

+ \frac{5}{2} g_{d} g_\epsilon \tr\left(Y_\nu^{\dagger} Y_\nu \right) Y_\nu

+ \frac{5}{8} g_\epsilon^{2} \tr\left(Y_\nu^{\dagger} Y_\nu \right) Y_\nu

+ \frac{15}{8} g_2^{2} \tr\left(Y_\nu^{\dagger} Y_\nu \right) Y_\nu

+ \frac{35}{24} g_{Y}^{4} Y_\nu

+ \frac{187}{12} g_{Y}^{2} g_{d}^{2} Y_\nu

+ \frac{35}{12} g_{Y}^{2} g_\epsilon^{2} Y_\nu

+ 65 g_{d}^{4} Y_\nu

+ \frac{799}{12} g_{d}^{2} g_\epsilon^{2} Y_\nu

+ 21 g_{Y}^{2} g_{d} g_\epsilon Y_\nu

+ 21 g_{d} g_\epsilon^{3} Y_\nu

+ \frac{9}{4} g_2^{2} g_{d} g_\epsilon Y_\nu

+ \frac{253}{3} g_{d}^{3} g_\epsilon Y_\nu

+ \frac{35}{24} g_\epsilon^{4} Y_\nu

-  \frac{9}{4} g_2^{2} g_{Y}^{2} Y_\nu

+ \frac{27}{4} g_2^{2} g_{d}^{2} Y_\nu

-  \frac{9}{4} g_2^{2} g_\epsilon^{2} Y_\nu

-  \frac{23}{4} g_2^{4} Y_\nu
\end{autobreak}
\end{align*}
\begin{align*}
\begin{autobreak}
\beta^{(1)}(Y_N) =

+ Y_\nu^{\trans} Y_\nu^{*} Y_N

+ Y_N Y_\nu^{\dagger} Y_\nu

+ 4 Y_N Y_N^{*} Y_N

+ 2 \tr\left(Y_N^{*} Y_N \right) Y_N

- 6 g_{d}^{2} Y_N
\end{autobreak}
\end{align*}
\begin{align*}
\begin{autobreak}
\beta^{(2)}(Y_N) =

-  \frac{1}{4} Y_\nu^{\trans} Y_\nu^{*} Y_\nu^{\trans} Y_\nu^{*} Y_N

+ 4 Y_\nu^{\trans} Y_\nu^{*} Y_N Y_\nu^{\dagger} Y_\nu

-  \frac{1}{4} Y_N Y_\nu^{\dagger} Y_\nu Y_\nu^{\dagger} Y_\nu

-  Y_N Y_\nu^{\dagger} Y_\nu Y_N^{*} Y_N

-  Y_N Y_N^{*} Y_\nu^{\trans} Y_\nu^{*} Y_N

+ 28 Y_N Y_N^{*} Y_N Y_N^{*} Y_N

-  \frac{9}{2} \tr\left(Y_u^{\dagger} Y_u \right) Y_\nu^{\trans} Y_\nu^{*} Y_N

-  \frac{9}{2} \tr\left(Y_u^{\dagger} Y_u \right) Y_N Y_\nu^{\dagger} Y_\nu

-  \frac{3}{2} \tr\left(Y_\nu^{\dagger} Y_\nu \right) Y_\nu^{\trans} Y_\nu^{*} Y_N

-  \frac{3}{2} \tr\left(Y_\nu^{\dagger} Y_\nu \right) Y_N Y_\nu^{\dagger} Y_\nu

- 6 \tr\left(Y_\nu^{\dagger} Y_\nu Y_N^{*} Y_N \right) Y_N

- 12 \tr\left(Y_N^{*} Y_N Y_N^{*} Y_N \right) Y_N

- 12 \tr\left(Y_N^{*} Y_N \right) Y_N Y_N^{*} Y_N

- 4 \lambda_3 Y_\nu^{\trans} Y_\nu^{*} Y_N

- 4 \lambda_3 Y_N Y_\nu^{\dagger} Y_\nu

- 32 \lambda_2 Y_N Y_N^{*} Y_N

+ 4 \lambda_2^{2} Y_N

+ \lambda_3^{2} Y_N

+ \frac{17}{8} g_{Y}^{2} Y_\nu^{\trans} Y_\nu^{*} Y_N

- 16 g_{d}^{2} Y_\nu^{\trans} Y_\nu^{*} Y_N

-  \frac{13}{2} g_{d} g_\epsilon Y_\nu^{\trans} Y_\nu^{*} Y_N

+ \frac{17}{8} g_\epsilon^{2} Y_\nu^{\trans} Y_\nu^{*} Y_N

+ \frac{51}{8} g_2^{2} Y_\nu^{\trans} Y_\nu^{*} Y_N

+ \frac{17}{8} g_{Y}^{2} Y_N Y_\nu^{\dagger} Y_\nu

- 16 g_{d}^{2} Y_N Y_\nu^{\dagger} Y_\nu

-  \frac{13}{2} g_{d} g_\epsilon Y_N Y_\nu^{\dagger} Y_\nu

+ \frac{17}{8} g_\epsilon^{2} Y_N Y_\nu^{\dagger} Y_\nu

+ \frac{51}{8} g_2^{2} Y_N Y_\nu^{\dagger} Y_\nu

+ 176 g_{d}^{2} Y_N Y_N^{*} Y_N

+ 10 g_{d}^{2} \tr\left(Y_N^{*} Y_N \right) Y_N

- 127 g_{d}^{4} Y_N

-  \frac{35}{6} g_{d}^{2} g_\epsilon^{2} Y_N

-  \frac{32}{3} g_{d}^{3} g_\epsilon Y_N
\end{autobreak}
\end{align*}

\subsection{Leptoquark $S_3$\label{app:LQRGE}}

In this model the 2-loop Yukawa RGEs depend also on the dimensionless BSM Higgs potential parameters $\lambda_{S_3}$ and $\lambda_{HS_3}$, which we define as
\begin{align}
    \left.\mathcal{V}\right|_{S_3} = \frac{1}{2} m_{S_3}^2 \text{Tr}\left[S_3^\dagger S_3\right] + \frac{1}{8} \lambda_{S_3} \left( \text{Tr}\left[S_3^\dagger S_3\right] \right)^2 + \frac{1}{2} \lambda_{HS_3}  H^\dagger H \, \text{Tr}\left[S_3^\dagger S_3\right]. \label{eq:S3_potential}
\end{align}

We have checked that these couplings do not influence predictions for the gauge and Yukawa sector in a significant manner. As such, they are set to 0 in our numerical analysis.

Note that there was an error in the normalization of $Y_{\textrm{LQ}}$ in the RGEs given in Ref.\cite{Kowalska:2020gie}.
Below we present the correct expressions.

\subsubsection{Gauge sector}
\begin{align*}
\begin{autobreak}
\beta^{(1)}(g_{Y}) =\frac{43}{6} g_{Y}^{3}
\end{autobreak}
\end{align*}

\begin{align*}
\begin{autobreak}
\beta^{(2)}(g_{Y}) =

+ \frac{23}{2} g_{Y}^{5}

+ \frac{25}{2} g_2^{2} g_{Y}^{3}

+ 20 g_3^{2} g_{Y}^{3}

-  \frac{17}{6} g_{Y}^{3} \tr\left(Y_u^{\dagger} Y_u \right)

- 5 g_{Y}^{3} \tr\left(Y_{\textrm{LQ}}^{\dagger} Y_{\textrm{LQ}} \right)
\end{autobreak}
\end{align*}

\begin{align*}
\begin{autobreak}
\beta^{(1)}(g_2) =- \frac{7}{6} g_2^{3}
\end{autobreak}
\end{align*}

\begin{align*}
\begin{autobreak}
\beta^{(2)}(g_2) =

+ \frac{25}{6} g_2^{3} g_{Y}^{2}

+ \frac{371}{6} g_2^{5}

+ 44 g_2^{3} g_3^{2}

-  \frac{3}{2} g_2^{3} \tr\left(Y_u^{\dagger} Y_u \right)

- 9 g_2^{3} \tr\left(Y_{\textrm{LQ}}^{\dagger} Y_{\textrm{LQ}} \right)
\end{autobreak}
\end{align*}

\begin{align*}
\begin{autobreak}
\beta^{(1)}(g_3) =- \frac{13}{2} g_3^{3}
\end{autobreak}
\end{align*}

\begin{align*}
\begin{autobreak}
\beta^{(2)}(g_3) =

+ \frac{33}{2} g_2^{2} g_3^{3}

+ \frac{5}{2} g_3^{3} g_{Y}^{2}

- 15 g_3^{5}

- 2 g_3^{3} \tr\left(Y_u^{\dagger} Y_u \right)

- 3 g_3^{3} \tr\left(Y_{\textrm{LQ}}^{\dagger} Y_{\textrm{LQ}} \right)
\end{autobreak}
\end{align*}

\subsubsection{Yukawa sector}

\begin{align*}
\begin{autobreak}
\beta^{(1)}(Y_u) =

+ \frac{3}{2} Y_u Y_u^{\dagger} Y_u

+ \frac{3}{2} Y_{\textrm{LQ}}^{\dagger} Y_{\textrm{LQ}} Y_u

+ 3 \tr\left(Y_u^{\dagger} Y_u \right) Y_u

-  \frac{17}{12} g_{Y}^{2} Y_u

-  \frac{9}{4} g_2^{2} Y_u

- 8 g_3^{2} Y_u
\end{autobreak}
\end{align*}

\begin{align*}
\begin{autobreak}
\beta^{(2)}(Y_u) =

+ \frac{3}{2} Y_u Y_u^{\dagger} Y_u Y_u^{\dagger} Y_u

-  \frac{3}{4} Y_u Y_u^{\dagger} Y_{\textrm{LQ}}^{\dagger} Y_{\textrm{LQ}} Y_u

-  \frac{27}{8} Y_{\textrm{LQ}}^{\dagger} Y_{\textrm{LQ}} Y_{\textrm{LQ}}^{\dagger} Y_{\textrm{LQ}} Y_u

-  \frac{27}{4} \tr\left(Y_u^{\dagger} Y_u Y_u^{\dagger} Y_u \right) Y_u

-  \frac{27}{4} \tr\left(Y_u^{\dagger} Y_u \right) Y_u Y_u^{\dagger} Y_u

-  \frac{27}{4} \tr\left(Y_u^{\dagger} Y_{\textrm{LQ}}^{\dagger} Y_{\textrm{LQ}} Y_u \right) Y_u

-  \frac{9}{2} \tr\left(Y_{\textrm{LQ}}^{\dagger} Y_{\textrm{LQ}} \right) Y_{\textrm{LQ}}^{\dagger} Y_{\textrm{LQ}} Y_u

- 12 \lambda Y_u Y_u^{\dagger} Y_u

- 6 \lambda_{HS_3} Y_{\textrm{LQ}}^{\dagger} Y_{\textrm{LQ}} Y_u

+ 6 \lambda^{2} Y_u

+ \frac{9}{2} \lambda_{HS_3}^{2} Y_u

+ \frac{223}{48} g_{Y}^{2} Y_u Y_u^{\dagger} Y_u

+ \frac{135}{16} g_2^{2} Y_u Y_u^{\dagger} Y_u

+ 16 g_3^{2} Y_u Y_u^{\dagger} Y_u

+ \frac{43}{6} g_{Y}^{2} Y_{\textrm{LQ}}^{\dagger} Y_{\textrm{LQ}} Y_u

+ \frac{45}{2} g_2^{2} Y_{\textrm{LQ}}^{\dagger} Y_{\textrm{LQ}} Y_u

+ 11 g_3^{2} Y_{\textrm{LQ}}^{\dagger} Y_{\textrm{LQ}} Y_u

+ \frac{85}{24} g_{Y}^{2} \tr\left(Y_u^{\dagger} Y_u \right) Y_u

+ \frac{45}{8} g_2^{2} \tr\left(Y_u^{\dagger} Y_u \right) Y_u

+ 20 g_3^{2} \tr\left(Y_u^{\dagger} Y_u \right) Y_u

+ \frac{449}{72} g_{Y}^{4} Y_u

-  \frac{3}{4} g_2^{2} g_{Y}^{2} Y_u

+ \frac{1}{4} g_2^{4} Y_u

+ 9 g_2^{2} g_3^{2} Y_u

+ \frac{19}{9} g_3^{2} g_{Y}^{2} Y_u

-  \frac{302}{3} g_3^{4} Y_u
\end{autobreak}
\end{align*}

\begin{align*}
\begin{autobreak}
\beta^{(1)}(Y_{\textrm{LQ}}) =

+ \frac{1}{2} Y_{\textrm{LQ}} Y_u Y_u^{\dagger}

+ 6 Y_{\textrm{LQ}} Y_{\textrm{LQ}}^{\dagger} Y_{\textrm{LQ}}

+ 2 \tr\left(Y_{\textrm{LQ}}^{\dagger} Y_{\textrm{LQ}} \right) Y_{\textrm{LQ}}

-  \frac{5}{6} g_{Y}^{2} Y_{\textrm{LQ}}

-  \frac{9}{2} g_2^{2} Y_{\textrm{LQ}}

- 4 g_3^{2} Y_{\textrm{LQ}}
\end{autobreak}
\end{align*}

\begin{align*}
\begin{autobreak}
\beta^{(2)}(Y_{\textrm{LQ}}) =

-  \frac{1}{4} Y_{\textrm{LQ}} Y_u Y_u^{\dagger} Y_u Y_u^{\dagger}

-  \frac{9}{8} Y_{\textrm{LQ}} Y_u Y_u^{\dagger} Y_{\textrm{LQ}}^{\dagger} Y_{\textrm{LQ}}

+ \frac{13}{4} Y_{\textrm{LQ}} Y_{\textrm{LQ}}^{\dagger} Y_{\textrm{LQ}} Y_{\textrm{LQ}}^{\dagger} Y_{\textrm{LQ}}

-  \frac{9}{4} \tr\left(Y_u^{\dagger} Y_u \right) Y_{\textrm{LQ}} Y_u Y_u^{\dagger}

-  \frac{3}{2} \tr\left(Y_u^{\dagger} Y_{\textrm{LQ}}^{\dagger} Y_{\textrm{LQ}} Y_u \right) Y_{\textrm{LQ}}

- 18 \tr\left(Y_{\textrm{LQ}}^{\dagger} Y_{\textrm{LQ}} Y_{\textrm{LQ}}^{\dagger} Y_{\textrm{LQ}} \right) Y_{\textrm{LQ}}

- 18 \tr\left(Y_{\textrm{LQ}}^{\dagger} Y_{\textrm{LQ}} \right) Y_{\textrm{LQ}} Y_{\textrm{LQ}}^{\dagger} Y_{\textrm{LQ}}

- 2 \lambda_{HS_3} Y_{\textrm{LQ}} Y_u Y_u^{\dagger}

- 24 \lambda_{S_3} Y_{\textrm{LQ}} Y_{\textrm{LQ}}^{\dagger} Y_{\textrm{LQ}}

+ 5 \lambda_{S_3}^{2} Y_{\textrm{LQ}}

+ \lambda_{HS_3}^{2} Y_{\textrm{LQ}}

+ \frac{427}{144} g_{Y}^{2} Y_{\textrm{LQ}} Y_u Y_u^{\dagger}

+ \frac{33}{16} g_2^{2} Y_{\textrm{LQ}} Y_u Y_u^{\dagger}

-  \frac{8}{3} g_3^{2} Y_{\textrm{LQ}} Y_u Y_u^{\dagger}

+ \frac{23}{3} g_{Y}^{2} Y_{\textrm{LQ}} Y_{\textrm{LQ}}^{\dagger} Y_{\textrm{LQ}}

+ 96 g_2^{2} Y_{\textrm{LQ}} Y_{\textrm{LQ}}^{\dagger} Y_{\textrm{LQ}}

+ 62 g_3^{2} Y_{\textrm{LQ}} Y_{\textrm{LQ}}^{\dagger} Y_{\textrm{LQ}}

+ \frac{25}{18} g_{Y}^{2} \tr\left(Y_{\textrm{LQ}}^{\dagger} Y_{\textrm{LQ}} \right) Y_{\textrm{LQ}}

+ \frac{15}{2} g_2^{2} \tr\left(Y_{\textrm{LQ}}^{\dagger} Y_{\textrm{LQ}} \right) Y_{\textrm{LQ}}

+ \frac{20}{3} g_3^{2} \tr\left(Y_{\textrm{LQ}}^{\dagger} Y_{\textrm{LQ}} \right) Y_{\textrm{LQ}}

+ \frac{599}{144} g_{Y}^{4} Y_{\textrm{LQ}}

-  \frac{23}{24} g_2^{2} g_{Y}^{2} Y_{\textrm{LQ}}

-  \frac{173}{16} g_2^{4} Y_{\textrm{LQ}}

- 31 g_2^{2} g_3^{2} Y_{\textrm{LQ}}

-  \frac{1}{9} g_3^{2} g_{Y}^{2} Y_{\textrm{LQ}}

-  \frac{55}{3} g_3^{4} Y_{\textrm{LQ}}
\end{autobreak}
\end{align*}

\clearpage
\printbibliography[title={References}]
\addcontentsline{toc}{section}{References}

\end{document}